\documentclass[aps,onecolumn]{revtex4-2}
\usepackage{amssymb,amsfonts,amsmath,bm}
\usepackage{color}
\usepackage{graphicx}
\usepackage{cancel}
\usepackage{subfigure}
\usepackage{amssymb, latexsym, amsmath}
 \usepackage[T2A]{fontenc}

\begin{document}

\title{Pattern formation in nonlinear dynamics of nematic \\
liquid crystals above the flexoelectric instability threshold}

\author{E.S. Pikina$^{1}$, A.R. Muratov$^{2}$, E.I. Kats$^1$, V.V. Lebedev$^{1,3}$.}

\affiliation{$^1$ Landau Institute for Theoretical Physics, RAS, \\
	142432, Chernogolovka, Moscow region, Russia, \\
	$^2$ Institute for Oil and Gas Research, RAS, 119917, Gubkina 3, Moscow, Russia,
	$^3$ NRU Higher School of Economics, \\
	101000, Myasnitskaya 20, Moscow, Russia. }

\begin{abstract}

For many decades, researchers have been studying various types of electro-hydrodynamic instabilities in liquid crystals. A significant amount of experimental data has been collected, however, the theoretical interpretations of the results typically rely on linear analysis. In response to this limitation, we investigate the nonlinear stage of the flexoelectric instability in nematics, focusing on liquid crystals with a negative anisotropy in their dielectric permittivity and electrical conductivity. We base our analysis on a comprehensive set of nonlinear electro-hydrodynamic equations for these nematics influenced by an external alternating electric field. The equations predict an instability that is driven by the flexoelectric effect. In order to examine the peculiarities of this phenomenon, we use a model that was proposed in our previous publications, Refs. \cite{PM23,PK24}, which allows us to perform numerical simulation of nonlinear dynamics. We examine patterns that are formed above the instability threshold. Through numerical simulations, we have identified static and dynamic patterns that occur over a timescale that is much longer than the period of the external electric field. The static patterns are one-dimensional structures and dynamic patterns are standing or traveling one-dimensional waves. The type of the realized pattern depends on the material and experimentally controlled parameters. We found that the standing waves are stable with respect to small transverse perturbations, whereas the propagating waves are unstable. We present a Ginzburg-Landau-like phenomenology that applies near the instability threshold. This approach allows us to rationalize our numerical findings with a few parameters.

\end{abstract}


\maketitle
\section{Introduction}
\label{sec:intro}

Physical systems that are driven far from equilibrium can exhibit spontaneous symmetry breaking, being organized into spatio-temporal patterns. The study of these phenomena has been a long-standing topic in condensed matter physics, investigated both experimentally and theoretically. In particular, attention has been paid to non-equilibrium pattern formation in liquid crystals, which has led to a vast number of publications on this topic. For those new to the field, we mention only a few key monographs \cite{PI91,CH92,GP93,BC94,BK96,KL03,OP05} and a relatively modern review \cite{ES16}, in which many relevant references can be found. Despite the numerous publications on pattern formation in liquid crystals, this phenomenon remains an active area of research. The interest in liquid crystals stems from their potential technological applications, as well as their role as an ideal testing ground for studying pattern formation phenomena. Additionally, liquid crystals represent a generic example of non-equilibrium dissipative processes, making them convenient for researchers.

In this work, we aim to investigate the nonlinear electro-hydrodynamics of nematic liquid crystals under the influence of an external alternating electric field. This is an ongoing area of research, and while this paper is not meant to be a review, we would like to outline some historical milestones that led to our current work. The history of this field started in the 1960s, with the first observations of the electro-hydrodynamic instability and theoretical understanding of the phenomenon, see Refs. \cite{WI63,HZ68,HE69}. The efforts led to the development of the so-called Standard Model (SM), which provides a detailed and quantitative description of a vast amount of experimental data, see Refs. \cite{v1,v2,v3,v4,v5,KN04,KP08,TE08,KD15,PK18}. Two key material parameters in the SM are the dielectric anisotropy, $\epsilon_a$, and the anisotropy of electrical conductivity, $\sigma_a$, which are differences of the components parallel and perpendicular to the director. It is convenient to classify all nematic materials into 4 classes, according to the signs of $\epsilon_a$ and $\sigma_a$, as $(++)$, $(+-)$, $(-+)$ and $(--)$ materials, see Ref. \cite{GP93}.

The SM predicts and describes reasonably well the linear stage of the electro-hydrodynamic instabilities for nematic materials of the first 3 classes, namely, $(++)$, $(+-)$, and $(-+)$. The matter is that the principle mechanism proposed first by Carr and Helfrich (see the pedagogical textbook description in Refs. \cite{PI91,CH92,GP93,BC94}) is the instability driven by the periodic director distortions out-off the initially planar alignment. In own turn the tilt via anisotropic conductivity $\sigma _a$ for sufficiently large voltage produces the positive feedback destabilizing the homogeneous and stationary nematic state. It is easy to see from this argument, that the mechanism, based on the competition electric charge and nematic director torques, works for the $(+ -)$ and $(- +)$ materials. In the $(+ +)$ materials due to the Fredericksz distortion of the uniform director alignment \cite{CH92,GP93}, the Carr-Helfrich mechanism is in a competition with Fredericksz transition. Evidently, in the $(- -)$ materials, the Carr-Helfrich mechanism does not lead to any instability, which is in contradiction with experimental data (see, e.g., the monograph \cite{BK96} and review paper \cite{ES16}). The contradiction is circumvented and eliminated by taking into account the flexoelectric effect, which is usually believed to be insignificant in the standard Carr-Helfrich approach. However, the intrinsic scaling of the flexoelectricity (its proportionality to the nematic director gradients) can cause physically relevant responses even for materials with relatively small flexoelectric coefficients. The generalization of the SM with flexoelectric effects taken into account, proposed in the works \cite{KP08,TE08,KP08} (and summarized in the review paper \cite{ES16}), called by the authors a non-standard model,  allows one to describe satisfactory the stationary textures observed in $(- -)$ materials.

However, it is not the end of the story. The matter is that the non-standard model predicts and describes only static patterns, that is time independent through the time scale much larger than the period  of the external alternating field, whereas experimentally moving or oscillating in time patterns are often observed, see Refs. \cite{v1,v2,v3,v4,v5,ES16,TC07}. For example, the authors of Ref. \cite{KP08} state that they were unable to derive from their linear analysis the Hopf bifurcation, observed in the their experimental studies \cite{TC07}. Our results contradicting this negative statement were obtained for the range of material and experimentally controlled parameters satisfying the condition  $\omega \gg {\sigma}/({\epsilon_0 \epsilon_\perp})$, see Ref. \cite{PK24}, not touched in the work \cite{KP08}.

Note that experimentally, it is very difficult (and practically impossible using only polarization microscopy methods) to distinguish between propagating and standing wave patterns. For example, standing waves will give images very similar to those shown in Fig. 14 of Ref. \cite{TC07}, where travelling waves are discussed.  Additional difficulties for the experimental identification of propagating and standing one-dimensional patterns are associated with the fact that for  one-dimensional patterns (rolls in the standard liquid-crystalline terminology) due to incompressibility condition the large time scale hydrodynamical  velocity component along the   traveling waves propagation direction is  zero. It prevents an identification of the travelling waves by using impurity particles (ink).

In turn, as we noted in Ref. \cite{PM23}, the starting point to investigate theoretically dynamic phenomena in liquid crystals is the complete set of the dynamic equations for all soft variables related to the general conservation laws and to symmetry breaking. In the considered case of sufficiently low electric conductivity $\sigma $, the electromagnetic degrees of freedom should be included into consideration as well. Contrary, in the works \cite{TK95,DT96,Ab23} a reduced system of equations was used. It is potentially dangerous, as some effects may be overlooked in this approach. Note also that, in contrast to the abstract Ginzburg-Landau approach used in the work \cite{ZI91}, the phenomenology presented in our work is constructed in accordance with the results obtained through numerical modeling.

Another pass to explain experimentally observed dynamic patterns is to add new ingredients into the non-standard model. For example, it was suggested in Ref. \cite{AE22} to combine the non-standard model with the so-called weak electrolyte model. The price of extending the model is the increase in the number of parameters (charge densities, anisotropic mobilities and recombination rates of ionic species), usually poorly known. The complete theoretical description of nonlinear stages of pattern evolution within the framework of such an extended model is hopeless. Indeed, one has to solve a set of 10 nonlinear and nonlocal integro-differential equations, what is computationally prohibitive without some dramatic (and not well controlled by small parameters) simplifications of the system of the equations. As a result, there has not yet been a complete solution, even in the case of a linear approximation.

Recently new and important results were obtained in the realm of electro-hydrodynamic instabilities in liquid crystals, see Refs. \cite{LA1,LA2,LA3,dierking1,dierking2,dierking5,aya,new-26,new-1-26}. Namely, three-dimensionally localized excitations (solitons, directons, bullets, and skyrmions) have been reported as the results of experimental and numerical studies of hydro-electrodynamic instabilities in nematics and cholesterics. However, a complete theoretical description of the phenomena presented in these works is still absent and is a subject of subsequent research. This is one of potential applications of the theoretical framework we proposed in our paper.

We decided to simplify the model instead of an uncontrolled simplification of the system of equations. In Ref. \cite{PM23} we proposed a simplified model (termed as the minimal model in \cite{PK24}) but still reach enough to keep all the essential physics of dynamic bifurcations, i.e., without throwing "the baby out of the bathwater". Restricting ourselves to this model and $(- -)$ materials, we solve numerically the complete system of the nonlinear dynamic equations of a nematic, previously obtained by our group, see Ref. \cite{PM23}. Some simulations made using an extension of the model have confirmed its validity.

In our simulations we use the standard setup where a plane-parallel film of a nematic is placed in a capacitor with an alternating voltage applied to its electrodes. Below the threshold of the flexoelectric instability the director field $\bm n$ is homogeneous, $\bm n$ is oriented parallel to the plates. The instability occurs for a mode with a finite wave vector $\bm q_c$, oriented parallel to the capacitor plates. We found in Ref. \cite{PK24} that characteristics of the flexoelectric instability are insensitive to boundary conditions on the plates provided $q_c d \gg 1$, where $d$ is the thickness of the film. Then only thin boundary layers feel the boundary conditions whereas the bulk does not. That is why in the linear approximation the set of  partial differential equations can be reduced to a set of ordinary differential equations for spacial Fourier harmonics.

Patterns that are formed above the instability threshold can only be studied by using nonlinear equations. At the condition $q_c d \gg 1$ a $z$-dependence of the fields can be neglected, where $z$ is the coordinate in the direction perpendicular to the capacitor plates. Thus the problem is reduced to a two-dimensional one, where all fields are functions of the coordinates $x$ and $y$. To solve the problem we use numerical simulations with the periodic boundary conditions in $x-y$ plane in the box determined by the components of the critical wave vector $\bm q_c$. It enables us to catch the critical mode and its dynamics. Note that the nonlinearity involves into the game other degrees of freedom influencing the state above the instability threshold.

Let stress, that we study the slow dynamics, that is the pattern dynamics on time scales much larger than the period $T= 2\pi/\omega$ of the external alternating electric field, whereas oscillations of the pattern with the period $T$ are omnipresent. From this point of view we use the term static solution for the solution when its envelope doesn't change through this long time scale. One can expect a variety of possible solutions, including one-dimensional  static patterns or standing and traveling waves, as well as two-dimensional space-modulated and time-oscillating textures. Their complete classification only makes sense as a classification based on observations rather than on theoretical games or numerical simulations with a few dozen parameters. The linear stability analysis, performed in Ref. \cite{PK24}, assumes that the inhomogeneous perturbations are infinitely small. To describe the patterns above the bifurcation threshold we have to consider finite amplitudes of the inhomogeneous components of the fields. Therefore, nonlinear terms have to be taken into account in the analysis in order to distinguish between different possible structures. This is the goal of this work.

In our numeric simulations we focus on the range of controlling behavior parameters driving the system to the vicinity of the instability threshold. First we examine one-dimensional solutions. For such patterns, the nonlinear electro-hydrodynamic equations can be written in a compact form. Solutions of such equations can be found quite simply and quickly within the standard codes of Wolfram Mathematics. The next step is an investigation of stability of such one-dimensional patterns under two-dimensional perturbations. The analysis shows that the static one-dimensional solutions are stable. Surprisingly, only one-dimensional standing waves are stable with respect to small transverse perturbations, whereas propagating one-dimensional waves are unstable with respect to such perturbations. The conclusions are made for the sets of parameters examined in our numerical simulations and can be corrected for other sets.

To rationalize all obtained regimes of dynamic patterns, we formulate Ginzburg-Landau-like phenomenology, based on our numerical results. The phenomenology describes all the patterns observed numerically in the weakly non-linear regime in terms of a few parameters. Potentially, it can be used for examining such non-linear objects as solitons.

Our paper is organized as follows. In the next section \ref{sec:nonlinear} we formulate the complete set
of the non-linear electro-hydrodynamic equations to be solved, and their simplifications used in our simulations. In Sec. \ref{sec:one-dim} the one-dimensional numerical scheme is formulated and the results of the simulations conducted using this approach are presented. In Sec. \ref{sec:two-dim} we study stability of the found in section \ref{sec:one-dim} one-dimensional patterns with respect to transverse perturbations. In Sec. \ref{sec:phen} we formulate the Ginzburg-Landau like phenomenology valid in the vicinity of the instability threshold. We summarize the findings of this work, some open challenging problems, and the next steps required to address them in the concluding section \ref{sec:conclusion}. Some technical details of the derivation of the equations and the approximations made are presented in Appendix.

\section{Nonlinear electro-hydrodynamics of nematics}
\label{sec:nonlinear}

Here we discuss the nonlinear equations of nematics in a form convenient for our purpose, that is to describe the flexoelectric instability in $(- -)$ materials. The complete system of nonlinear equations of nematics in the external alternating electric field was derived in our paper \cite{PM23}. We use some technical simplifications to obtain a more compact system of equations that still keeps all essential physics of the electro-hydrodynamic phenomena in nematic liquid crystals.

We assume that the Mach number is small, so we can consider the nematic to be incompressible. This means that its mass density, $\rho$, is uniform, and the velocity of the nematic, $\bm{v}$, is non-divergent, with $\nabla\cdot\bm{v}=0$. We assume Reynolds number to be small as well. In this case the nonlinear hydrodynamic interaction is negligible. If the thermal conductivity tensor of the nematic is large then its temperature $T$ is homogeneous. If the thermal conductivity tensor is small then the specific entropy $S/\rho$ is homogeneous. Both limit cases enable one to exclude temperature (entropy) from the consideration. We use mainly the single-constant approximation $K=K_1=K_2=K_3$ for Frank energy. For a few set of the model parameters we check that all obtained results are qualitatively correct for the non-equal Frank modules (for details see Appendix A).

The non-homogeneity of the director field $\bm n$ generates the following electrical polarization vector of the nematic \cite{Mey69}
\begin{eqnarray}
\bm P_{fl} \, =  \, (e_1 - e_2)\, \bm n (\nabla \bm n) \, + \, e_2 \, \partial_k (n_k \bm n  ) \,  . \
\label{weakc7fl}
\end{eqnarray}
Then the electric displacement field $\bm D$ is written as the sum
\begin{equation}
D_i = \epsilon_0\, \epsilon_{ik}E_k +  P_{fl, i}\, , \
\label{displ}
\end{equation}
where the permittivity tensor is
\begin{equation}
\epsilon_{ik}=\epsilon_\perp \delta^{\perp}_{ik}
+\epsilon_\parallel n_i n_k, \quad
\delta_{ik}^\perp\equiv \delta_{ik}-n_in_k \, . \
\label{epsilonpp}
\end{equation}
The dielectric anisotropy is $\epsilon_a=\epsilon_\parallel-\epsilon_\perp$.

Following the minimal model assumptions we include into the consideration only one flexoelectric coefficient $\zeta=e_1$ (regarding $e_2\to 0$), only a single hydrodynamic viscosity coefficient $\eta$ (that is the isotropic shear viscosity). Besides we take the reactive kinetic coefficient $\lambda$ to be equal to unity that corresponds to the ideal rod-like molecules.  Note that for a qualitative analysis of all possible bifurcation types and the resulting patterns, the minimal model is not so bad. For example, the flexoelectric coefficient $e_2$ is related to a small dipole-like deformation of the director, whereas the coefficient $e_1$ takes into account both dipole and quadrupole deformations and is larger in typical nematics. Considering only a single hydrodynamic viscosity coefficient (in addition to the director friction coefficient) is also entirely justified for a qualitative analysis of possible bifurcations and the resulting patterns. The fact is that the hydrodynamic motion of the fluid plays a relatively minor quantitative role, because the critical mode is mainly determined by fluctuations of the director and the electric field potential. However, we may not simply discard the hydrodynamic motion, since dynamic Hopf bifurcations are only possible by taking into account hydrodynamic degrees of freedom. Similar qualitative arguments can be given also for the other simplifications of the minimal model.

With the simplifications we stay with the following nonlinear dynamic equation for director
\begin{eqnarray}
\partial_t n_i=
- \bm v \nabla n_i +n_k \delta_{ij}^\perp \partial_k v_j
+  \Xi_i^\perp/\gamma, \quad
\Xi_i^\perp = \delta^\perp_{ij}\Xi_j \, , \
\label{trun4}
\end{eqnarray}
where $\gamma$ is the rotational viscosity coefficient and
\begin{eqnarray}
\Xi_i =K \nabla^2 n_i
+ \epsilon_0\, (\epsilon_\parallel-\epsilon_\perp) E_i \bm n \bm E
+\zeta E_i (\nabla \bm n) -\zeta\partial_i(\bm n \bm E) \, , \
\label{trun6}
\end{eqnarray}
is minus variational derivative of the thermodynamic potential over $\bm n$. Since the director $\bm n$ is a unit vector, the equation (\ref{trun6}) is the equation for two degrees of freedom. Below we use the components $n_y,n_z$ as the parameters determining $\bm n$, then $n_x=(1-n_y^2-n_z^2)^{1/2}$.

The equation for the momentum density $\rho \bm v$ is its conservation law
\begin{eqnarray}
\rho \partial_t v_i =
- \nabla \left(K\nabla n_j \partial_i n_j\right)
- \partial_k(\Xi_i^\perp n_k)
- \partial_i P + \partial_k \left(D_k E_i\right)
+ \eta \nabla^2 v_i \, , \
\label{trun3}
\end{eqnarray}
where $P$ is pressure. To determine the pressure $P$, one should take the divergence of Eq. (\ref{trun3}) and use the condition $\nabla \bm{v} = 0$ to obtain a relation for $P$. In the case of equal elastic constants, the relation is
\begin{equation}
\nabla^2 P=
-\partial_k\partial_i \left(K\partial_k n_j \partial_i n_j\right)
-\partial_k\partial_i(\Xi_i^\perp n_k)
 + \partial_k\partial_i \left(D_k E_i\right) .
\label{trunp}
\end{equation}
The general equations for the pressure and velocity components for the case of different Frank elastic modules are given in Appendix A, see Eqs. (\ref{trun3K}) and (\ref{trunpK}).

We neglect effects related to the magnetic field, which are weak in nematics. Then it follows from the Maxwell equation $\partial_t \bm B +c \nabla\times \bm E=0$ that $\nabla\times \bm E=0$ that is the electric field is potential. Taking divergence of the other Maxwell equation $c\nabla\times \bm H=\partial_t \bm D + \bm j$ one finds
\begin{eqnarray}
\partial_t(\nabla \bm D)= - \partial_i(\sigma_{ik}E_k),
\label{trun2}
\end{eqnarray}
where we substituted the expression $j_i=\sigma_{ik} E_k$ for the current density of free charges. The components of the conductivity tensor are
\begin{eqnarray}
\sigma_{ik}=\sigma_\perp \delta_{ik}^\perp+
\sigma_\parallel n_i n_k \, .
\label{sigmacoeff}
\end{eqnarray}
Recall, that we consider the case where the conductivity anisotropy $\sigma_a=\sigma_\parallel - \sigma_\perp$ is negative. The system of equations (\ref{trun6},\ref{trun3},\ref{trun2}) constitute the closed system of dynamic equations for director $\bm n$, electric potential, and velocity $\bm v$, satisfying the incompressibility condition $\nabla \bm v=0$.

We split the electric field into two parts
\begin{equation}
\bm E = \bm E_0 \cos(\omega t)+ \bm E_1,
\nonumber
\end{equation}
where the first term on the right hand side with $\bm E_0$ directed along the $Z$-axis is the external electric field, and $\bm E_1$ is the electric field caused by polarization and free charges. We introduce the potential $\Phi$ of this induced field, $\bm E_1= -\nabla \Phi$. Then the displacement field (\ref{displ}) is written as
\begin{eqnarray}
D_{i} =  \epsilon_0\,\big( \epsilon_\perp \delta_{ik}^\perp +
\epsilon_\parallel n_i n_k \big) \big[E_0 \cos(\omega t) \delta_{kz} -
\partial_k \Phi \big] +  \zeta n_i \partial_k n_k \, .  \
\label{Dik}
\end{eqnarray}
Thus the equation (\ref{trun2}) appears to be the equation for the induced potential $\Phi$.

We examined numerically the nematodynamics above the flexoelectric instability, which occurs at a finite wave vector $q_c$. We consider the case where the thickness of the layer, $d$, is large compared to $\pi / q_c$, i.e., $q_c d\gg 1$. Then the derivatives with respect to the $z$-direction, $\partial_z$, in the dynamic equations are much smaller than the derivatives in the $x$- and $y$-directions, $\partial_x$ and $\partial_y$, respectively. Therefore, we can neglect the $\partial_z$ terms and stay with a two-dimensional problem. The incompressibility condition in this case is written as $\partial_x v_x+\partial_y v_y=0$. The simplifications allow us to perform Wolfram Mathematica numerical simulations of the nonlinear nematic dynamics in a reasonable time-frame.

However, one should be cautious in a such two-dimensional limit. The matter is that in this limit any memory of the boundary conditions (forcing the equilibrium direction of $\bm{n}$ along the $x$-axis) is lost. Therefore, if one completely neglects the boundary conditions, an additional (non-physical) symmetry arises with respect to the director rotation in the $x\,-\, y$ plane, that is a parasitic zero mode appears. When solving the dynamic equations, the presence of this zero mode leads to the accumulation of numerical errors. To overcome this problem, we introduce an additional term
\begin{equation}
\delta^\perp_{ik} A_k (\bm A \bm n),
\label{aaddition}
\end{equation}
into the right-hand side of Eq. (\ref{trun4}). Here $\bm A$ is a small vector directed along the $x$-axis. The term (\ref{aaddition}) restores the prevailed orientation of $\bm n$ along the $x$-axis.

\subsection{Avoiding non-locality of the equations}
\label{subsec:fictitious}

The equation (\ref{trun2}) for the potential $\Phi$ and the relation (\ref{trunp}) for pressure are non-local. It can be challenging to deal with this non-locality when solving the equations numerically. That is why we introduce fictitious dynamics in order to convert the non-local relations into local equations. Let us describe the way of this transformation.

One can treat the quantity $\varrho_e=\nabla\bm D$ as an independent variable, satisfying the equation
\begin{eqnarray}
\partial_t \varrho_e = -\nabla \bm j \, , \
\label{ro}
\end{eqnarray}
where $\bm j=\hat \sigma \bm E$ is the current density of free charges. The Eq. (\ref{ro}) is the local equation on variable $\varrho_e$. In turn, one should guarantee the relation $\varrho_e=\nabla \bm D$. It can be conveniently done by introducing a fictitious dynamic equation for the potential $\Phi$:
\begin{eqnarray}
{\mathcal{C}} \partial_t \Phi  + \nabla \big( \hat\epsilon \epsilon_0 (\bm E_0 - \nabla \Phi)
+ \bm P_{fl} \big) =  \varrho_e  \ ,
\label{potdyn}
\end{eqnarray}
which is a local equation. If the value of the parameter $\mathcal{C}$ is small enough,
\begin{eqnarray}
{\mathcal{C}} \omega \ll \, \epsilon_0\, \epsilon  \, q_c^2  \, , \
\label{potdyn1}
\end{eqnarray}
then Eq. (\ref{potdyn}) gives a solution close to $\varrho_e=\nabla\bm D$. Thus the nonlocal equation (\ref{trun2}) is replaced by two local dynamical equations (\ref{ro}) and (\ref{potdyn}).

However, there is an additional nuisance related to the potential $\Phi$. Since the equation (\ref{potdyn}) is invariant under the transformation $\Phi \to \Phi + \mathrm{const}$, there is a zero mode in $\Phi$ (a contribution independent of the coordinates), which can increase stochastically due to numerical errors. To suppress the effect one introduces an additional term with the factor $G_1$ to the right hand side of Eq. (\ref{potdyn}) to obtain
\begin{eqnarray}
{\mathcal{C}} \partial_t \Phi  =  -\,G_1 \Phi +  \varrho_e
- \nabla \big( \hat\epsilon \epsilon_0 (\bm E_0 - \nabla \Phi)+ \bm P_{fl} \big) \ ,
\label{potdyn2}
\end{eqnarray}
The constant $G_1$ should be much smaller than $\epsilon_0\, \epsilon  \, q_c^2$.

Analogously one can introduce fictitious dynamics for pressure $P$. We replace the relation (\ref{trunp}) by the following dynamic equation
\begin{equation}
{\mathcal{C}_P} \partial_t P= -\,G_P P + \nabla^2 P + \partial_k\partial_i \left(K\partial_k n_j \partial_i n_j\right)
+ \partial_k\partial_i(\Xi_i^\perp n_k) - \partial_k\partial_i \left(D_k E_i\right) \ , \
\label{trunp1}
\end{equation}
which is local unlike the relation (\ref{trunp}). We introduce here the term $-\,G_P P$ analogous to one in Eq. (\ref{potdyn2}). The goal is the same as before: to prevent a stochastic increase of the homogeneous component of pressure $P$. The general equation for the pressure for the case of different Frank elastic modules is given in Appendix A, see Eq. (\ref{trunpK}).

{By analogy with the equations for other fields, to prevent a stochastic increase (due to numerical errors) of the homogeneous component of velocity components $v_i$, we introduce the additional terms $\, - G_v\,v_i$ into the equations (\ref{trun3}), providing the zero mode suppression. One can say, that the extra terms in the equations reflect the residual $z$-dependence of the fields.  Doing so, we provided two-dimensional numerical solution of the system of dynamic nonlinear equations (\ref{trun4},\ref{trun3},\ref{ro},\ref{potdyn2},\ref{trunp1}).}

\section{One-dimensional numerical scheme}
\label{sec:one-dim}

To analyze the possible stable patterns that may exist above the instability threshold, we used a one-dimensional setup in which all fields depend on a single spatial coordinate. The approach allows one to quickly scan the character of nematodynamics for a variety of parameters. However, the results obtained within the framework of this approach need to be further examined, as they may be unstable under certain two-dimensional perturbations. We postpone this analysis to the next section.

First of all, the studied system of parabolic equations is so-to-speak "stiff" (or rigid), i.e. with very different characteristic times for essential variables. To solve it effectively it is desirable to exclude the fast modes from consideration. This can be done easily for the velocity, it is possible to omit the terms with the temporal derivative of velocity in corresponding equations. As a result equations for the velocity are transformed to one-time relations. Then, in the approach, all the fields depend on a variable $\xi$, which is a linear combination of the coordinates $x$ and $y$. The combination is determined by the critical wave vector $(q_{cx}, q_{cy})$ extracted from the linear analysis, see Ref. \cite{PK24}. It is the wave vector of the mode that becomes unstable at the point of instability. The variable $\xi$ is $\xi=\cos\alpha x +\sin\alpha y$ where $\alpha=\arctan(q_{cy}/q_{cx})$. We solve the equations for fields dependent of $\xi$ inside the interval of length $2\pi/q_c$ with periodic boundary conditions.

Note that the continuity equation looks like $\partial_{\xi}v_{\xi}=0$. This means, that the velocity in $\xi $ direction is equal to zero. Thus the velocity in $x,y$ plane is perpendicular to $\bm q_c$, we designate its value as $v_\chi$. In the situation the velocity does not enter the expression for pressure which is written as
\begin{equation}
P=-K(\partial_{\xi}n_j)^2-\Xi_{\xi}^{\perp}n_{\xi}+D_{\xi}E_{\xi} \ .
\end{equation}
Here, we introduce the components of the vectors along $\bm{q}_c$, as was done for velocity. In the one-dimensional setup, the pressure drops from the dynamic equations, since its gradient has the only component along the $\xi$ direction whereas the $\xi$ component of the velocity is zero.

The dependence of the fields on the single variable $\xi$ allows us to simplify significantly the system of equations (\ref{trun4}-\ref{trun3}). The equation (\ref{trun3}) for velocity acquires the form
\begin{eqnarray}
&&\eta\partial_{\xi}v_{\chi} =n_{\xi}(\Xi_{\chi}-n_{\chi}n_j\Xi_j)\ , \\ \nonumber
&&\eta\partial_{\xi}v_z =n_{\xi}(\Xi_z-n_z n_j\Xi_j)-D_{\xi} E_z \ \label{vzxi}
\end{eqnarray}
in components. Here we neglected the time derivatives and integrated once in $\xi$. Summing up the equations (\ref{vzxi}) we obtain
\begin{equation}
\eta n_j\partial_{\xi}v_j= -D_{\xi} E_z\ ,
\end{equation}
which is a compact relation.

The equation (\ref{trun4}) for director $\bm n$ in $1d$ acquires the form
\begin{eqnarray}
&& \partial_t\, n_y=n_{\xi}(\partial_{\xi}v_y-n_y n_j \partial_{\xi}v_j) +\Xi_y^{\perp} \ , \\
&& \partial_t\, n_z=n_{\xi}(\partial_{\xi}v_z-n_z n_j \partial_{\xi}v_j) +\Xi_z^{\perp} \ .
\nonumber
\end{eqnarray}
To prevent a stochastic rotation of director, we add the term (\ref{aaddition}) to the right hand side of the equations for $n_y,n_z$.
The equation (\ref{trun2}) remains in the original form. However, we use the fictitious dynamics for the potential $\Phi$, see Subsec. \ref{subsec:fictitious}. The values of additional parameters were chosen to satisfy the inequalities which are the applicability conditions of the fictitious dynamic, see Subsec. \ref{subsec:fictitious}.

\subsection{Phase diagram}
\label{subsec:phasedia}

The physical features of the flexoelectric instability depend on material parameters of the nematic. As it was demonstrated in Ref. \cite{PK24} the type of instability is determined mainly by the dimensionless combinations $\zeta/(K\epsilon_0|\Delta \epsilon|)^{1/2}$ and  $\eta/\gamma$, which we used in our linear stability analysis, as well as by the frequency $\omega$ of the external electric field. Therefore below we discuss our results in terms of these parameters.

In our case, the flexoelectric instability occurs at a finite wave vector. The wave vector $\bm{q}_c$ of the critical mode, which becomes unstable first when the external electric field is increased, can be found using a linear analysis, see Ref. \cite{PK24} for details. The results of the linear analysis show particularly, that the attenuation of the modes increases as the $z$-component of the wave vector, $q_z$, becomes non-zero. That is why the critical wave vector $\bm{q}_c$ has zero $z$-component.

\begin{figure}
\includegraphics[scale=0.4]{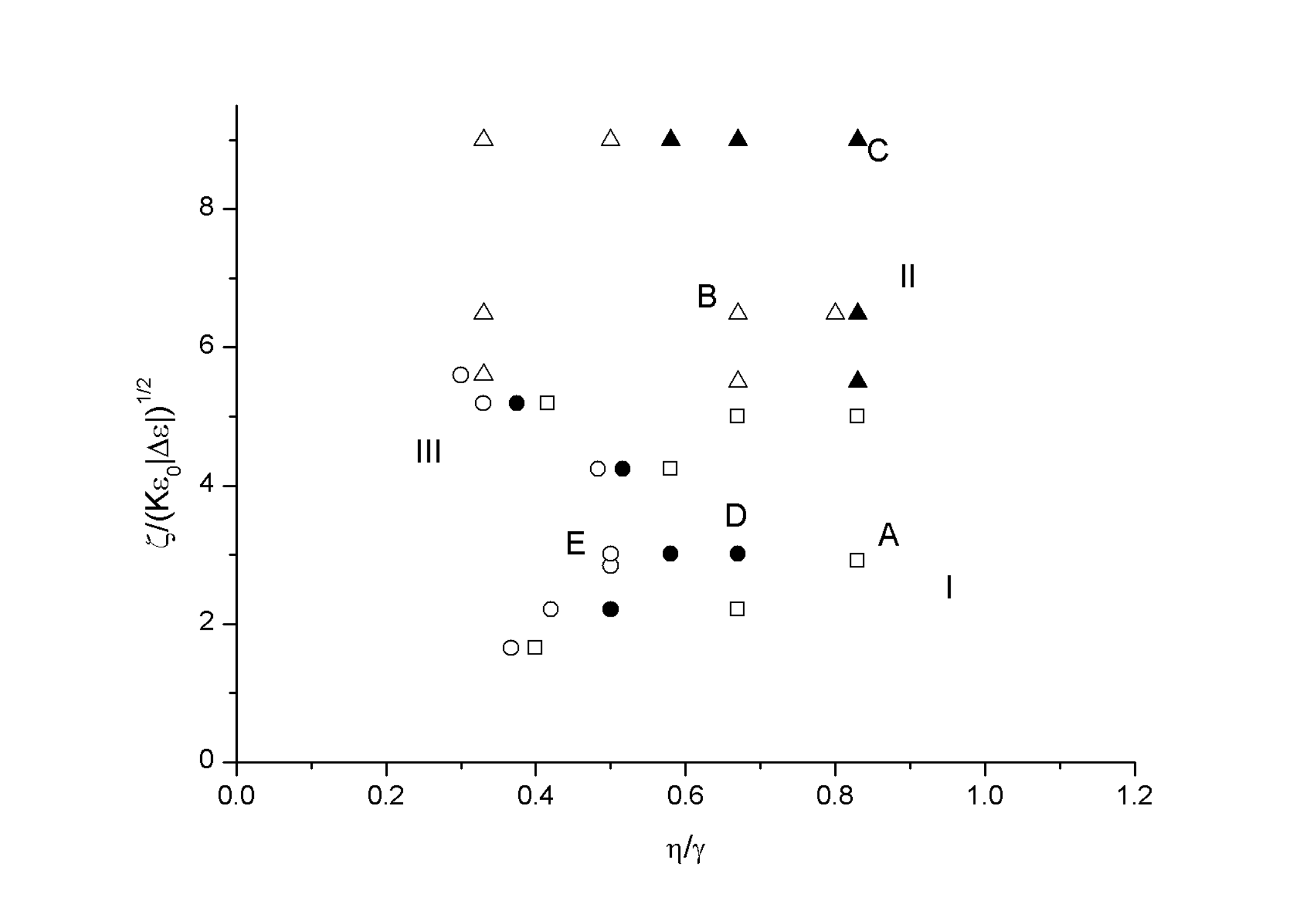}
\caption{Schematic "phase diagram" for critical excitations occurring at the instability threshold. There are three regions, I, II, and III, with different types of instability described in the text. Examined points in region I are marked by empty squares, examined points in region II are marked by triangles. In region III points are marked by filled circles for standing waves or empty circles for traveling waves. The letters marking the points indicate the parameter sets used for the numerical simulations, see Table \ref{tab:tab1}.}
\label{fig:pd}
\end{figure}

The "phase diagram" of the system, plotted in terms of the dimensionless parameters, $\eta/\gamma$ and $\zeta/(K\epsilon_0 |\Delta \epsilon|)^{1/2}$, was established in Ref. \cite{PK24}. The schematic view of the phase diagram is presented in Fig. \ref{fig:pd}. It illustrates the ranges of parameters associated with different types of instability, which are different in three regions, I, II, and III. The regions were identified using the linear instability analysis.

The region I corresponds to the critical mode with $q_{cx}=0$ (that is the critical mode is independent of $x$) and a real instability increment. Above the instability point one-dimensional patterns with layers parallel to the $x$-axis are formed in the region. The region II corresponds to the critical mode with both non-zero components of the wave vector $q_{cx},q_{cy}$ and a real instability increment. Above the instability point oblique one-dimensional patterns are formed in the region.  The region III corresponds to the critical mode with both non-zero components of the wave vector $q_{cx},q_{cy}$ and a complex instability increment. As demonstrated in Ref. \cite{PK24}, this feature is the result of the hybridization of two modes near the threshold, which are the electric potential mode and the director mode. In the region III above the instability threshold either standing or traveling wave can be formed.

The nonlinearity adds complexity to the phase diagram. At relatively small $\eta/\gamma$ the instability in the region II appears to be hard (with a finite jump, analogous to a first order phase transition). The examined points with such behavior are marked by empty triangles in Fig. \ref{fig:pd}. At growing $\eta/\gamma$ the instability approaches a soft (continuous) bifurcation. One can say, in the point B the behavior is close to a tricritical one, in terms of the thermodynamic phase transition semantics. At larger $\eta/\gamma$ another unstable mode with different wave vector comes into play. In this case first the instability leads to condensation of this second mode, the corresponding bifurcation is soft. However at the subsequent increase of the electric field the system returns by a jump to the basic critical mode. The examined points with such behavior are marked by filled triangles in Fig. \ref{fig:pd}.

The instability in the region I is soft. The examined points are marked by empty squares in Fig. \ref{fig:pd}. The instability in region III is soft as well. It can lead to both, traveling and standing waves. The examined points in the region are marked by empty circles for standing waves and by filled circles for traveling waves. We observe a transition from standing waves to traveling waves at increasing $\eta/\gamma$.

An important question concerns the dependence of the "phase diagram" on the frequency $\omega$ of the external field. From the linear analysis \cite{PK24} it follows that the critical field $E_{0c} \propto \sqrt{\omega}$ and the main wave vector of the one-dimensional pattern $q_c \propto \sqrt{\omega}$. Nonlinear analysis is of course much more complicated. Our simulations demonstrated that positions of the regions I, II and III only weakly depend on the frequency. In contrast, the boundaries between the sub-regions within the regions II and III depend on the frequency essentially. In the region II the increase of the frequency results in the shift of the boundary between the sub-regions of the first and second order transitions to the larger values of $\eta/\gamma$. In the region III the increase of the frequency reduces the stability region of the standing wave patterns.

To be more specific we present the magnitudes of the physical parameters for five points, A-D, in the phase diagram investigated in detail. The parameters are presented in Tab. \ref{tab:tab1}. In choosing the parameters we were guided by the values extracted from Refs. \cite{ES16,TE08,LA1,LA2,LA3}. The peculiarities of the instability in the points will be described below, see Sec. \ref{sec:one-dim}.

\begin{table}[!h]
\caption{Parameters of points studied in detail}
\begin{tabular*}{1.00\textwidth}{@{\extracolsep{\fill}}| c | c | c | c | c | c | c |}
\hline
Parameter          & Units        & A  &  B  & C  &  D &  E  \\ \hline
$\omega /(2\pi )$  & $s^{-1}$     & 500 & 500 &  400 & 400 & 400 \\ \hline
$K$                & $10^{-12}$ N & 4.  & 4. & 4. & 4.5 & 4.5 \\ \hline
$\gamma $          & Pa$\cdot$s   & \multicolumn{5}{ c |} {0.06} \\ \hline
$\eta $            & Pa$\cdot$s   & 0.05  & 0.04 & 0.05 & 0.04 & 0.03 \\ \hline
$\zeta$            &$10^{-11}/3$ C/m& 5.8 & 13 & 18. & 6.4 & 6. \\ \hline
$\epsilon_\perp $  & -            & \multicolumn{5}{ c |} {14} \\ \hline
$\Delta \epsilon $ & -            & \multicolumn{5}{ c |} {$-0.4\pi $} \\ \hline
$\sigma_\perp$     & $10^{-10}$ $\Omega^{-1} \cdot m^{-1}$     & 10  & 20 & 10 & 20 & 10 \\ \hline
$\Delta \sigma$    & $10^{-10}$ $\Omega^{-1} \cdot m^{-1}$     & -2 & -2 & -2 & -3 & -1 \\ \hline
$E_{0c}$           &$3\,10^4$ V/m & 141.3  & 65.9 & $\sim $ 41.; 47. & 125.2 & 125.2 \\ \hline
$q_{cx}$           & $\mu m^{-1}$ & 0. & 2.9 & 3.7 & 1.3 & 1.5 \\ \hline
$q_{cy}$           & $\mu m^{-1}$ & 3.1 & 5.5 & 8.3 & 3.3 & 3. \\ \hline
Region of diagram  & -            & I & II & II & III & III \\ \hline
Type of excitation & -            & stationary & stationary & stationary & standing & traveling \\
\hline
\end{tabular*}
\label{tab:tab1}
\end{table}

\subsection{Results of simulations}

The simplest case for analysis is the one with $q_{cx} = 0$, region I in Fig. \ref{fig:pd}. In this case the differential operator in the evolution equations is Hermitian, therefore it has only real eigenvalues. To study the solution of the problem we impose the periodic boundary conditions along $y$ direction with the period, obtained from the analysis of the linear system $2\cdot \pi/q_{cy}$. The stationary state of the solution can be determined by solving described system of the equations starting from small random non-zero initial conditions. If the external field amplitude $E_0$ is less than a threshold value $E_{0c}$, all fields are decreasing to zero values. If the electric field amplitude exceeds the critical value $E_{0c}$, the solution of the system of nonlinear equations stabilizes at a stationary state. This stationary state is identified as the appearance of stripes that are perpendicular to the $y$-axis.

\begin{figure}
\includegraphics[scale=0.5]{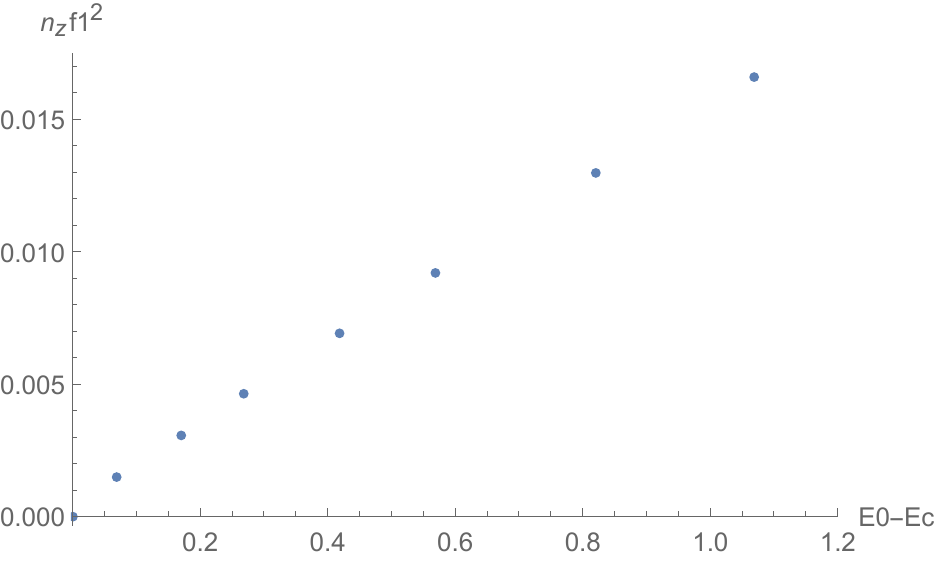}
\caption{The dependence of $n_z$ on the electric field for the excitation with $q_x=0$.
Parameters are presented in Table \ref{tab:tab1}, column A.}
\label{fig:0a}
\end{figure}

We examined the behavior of the amplitude of the critical mode on the electric field in the point A in Fig. \ref{fig:pd}, see Table \ref{tab:tab1}, column A. The blue circles in Figure \ref{fig:0a} present the dependence of the squared amplitude of main Fourier harmonic of $n_z$ in stripes on the amplitude of external field $E_0$ near the excitation threshold. As it follows from the figure the order parameter changes as square root of $\sqrt{E_0-E_{0c}}$, thus the transition is analogous to a second order phase transition.

Now we pass to the region II in Fig. \ref{fig:pd}, corresponding to the critical mode with non-zero $q_x$ and a real increment. The patterns appearing above the instability threshold are stationary stripes perpendicular to $\bm{e}_{\xi}$.

\begin{figure}
\includegraphics[scale=0.5]{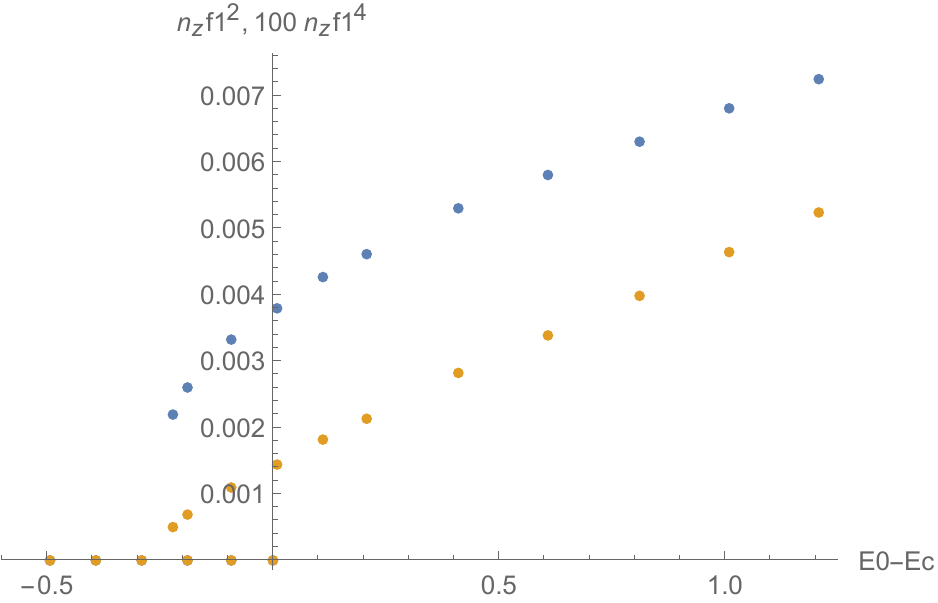}
\caption{The dependence of the value of $n_z$ on the electric field for the excitation with $q_x\ne 0$ and real increment. Blue points present the squared amplitude of the main harmonics of $n_z$, yellow points are $100\cdot n_z^4$. Parameters are presented in Table \ref{tab:tab1}, column B.}
\label{fig:re}
\end{figure}

We examined the behavior of the amplitude of the critical mode on the electric field in the point B in Fig. \ref{fig:pd}, see Table \ref{tab:tab1}, column B. Blue circles in Figure \ref{fig:re} show the obtained dependence of the squared amplitude of the main $n_z$ harmonic in the stripes near the excitation threshold on the amplitude of the external field $E_0$. In this case, the behavior of the order parameter resembles a first-order phase transition, with a rigid bifurcation. The yellow circles in Figure \ref{fig:re} represent the dependence of the fourth power of the amplitude on the external field, which is close to a linear relationship, indicating that the behavior is similar to a tricritical one. As can be seen from Figure \ref{fig:re}, in some range of external fields, both the homogeneous state and the state with the texture are stable, and the actual transition point lies within this region.

\begin{figure}
\includegraphics[height=4 cm]{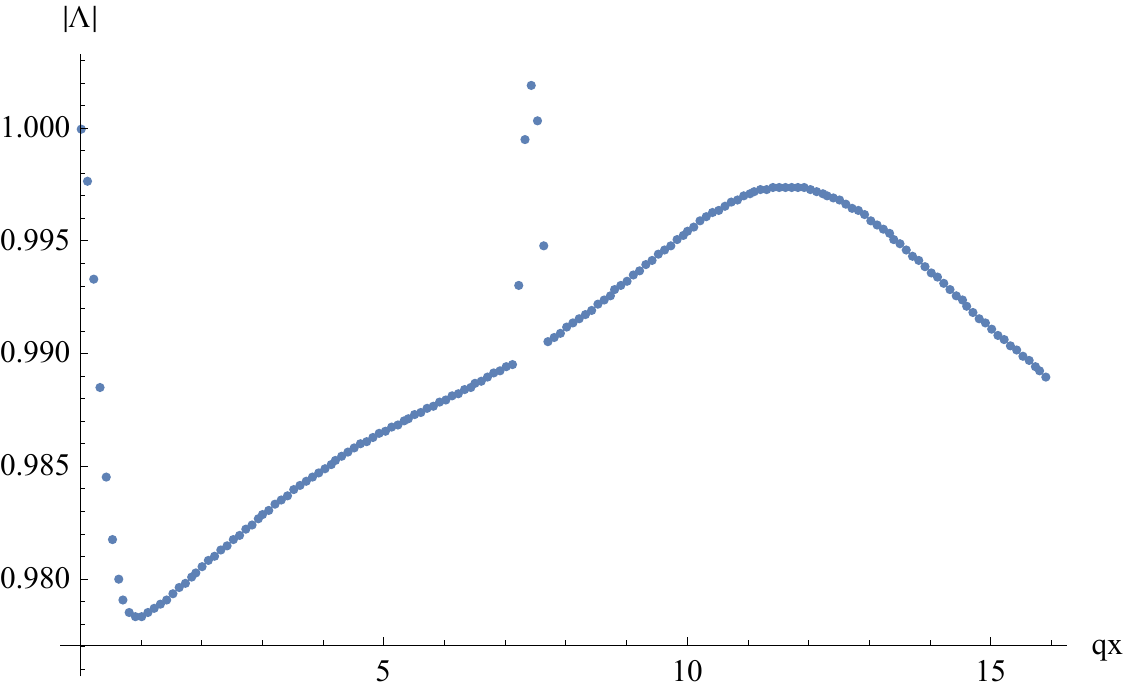}
\caption{The dependence of the main mode amplification factor on the wave vector $q_{\xi}$
for parameters set in Table \ref{tab:tab1}, column B.}
\label{fig:dispre}
\end{figure}

Figure \ref{fig:dispre} present the dependence of the factor $\Lambda$ of main mode on
the wave vector in the direction $e_{\xi}$ for the parameters in Table \ref{tab:tab1}, column B at the amplitude of external field close to the threshold. As it is seen in this case the parameter $\Lambda$ of main mode has two maxima: main maximum at wave vector
$q_1$ and an additional maximum at some wave vector $q_2 > q_1$. The dependence of the factor $\Lambda $ on the external field near $q_1$ and $q_2$ is different. The value of $\Lambda(q_1)$ strongly depends on $E_0$, whereas the value of $\Lambda(q_2)$ depends on $E_0$ weakly. The width of the maximum at $q_2$ is larger, than the width of the maximum at $q_1$. Then the exciting mode has the wave vector near $q_1$ and the corresponding bifurcation is a jump-like (''the first order'').

\begin{figure}
\includegraphics[height=4 cm]{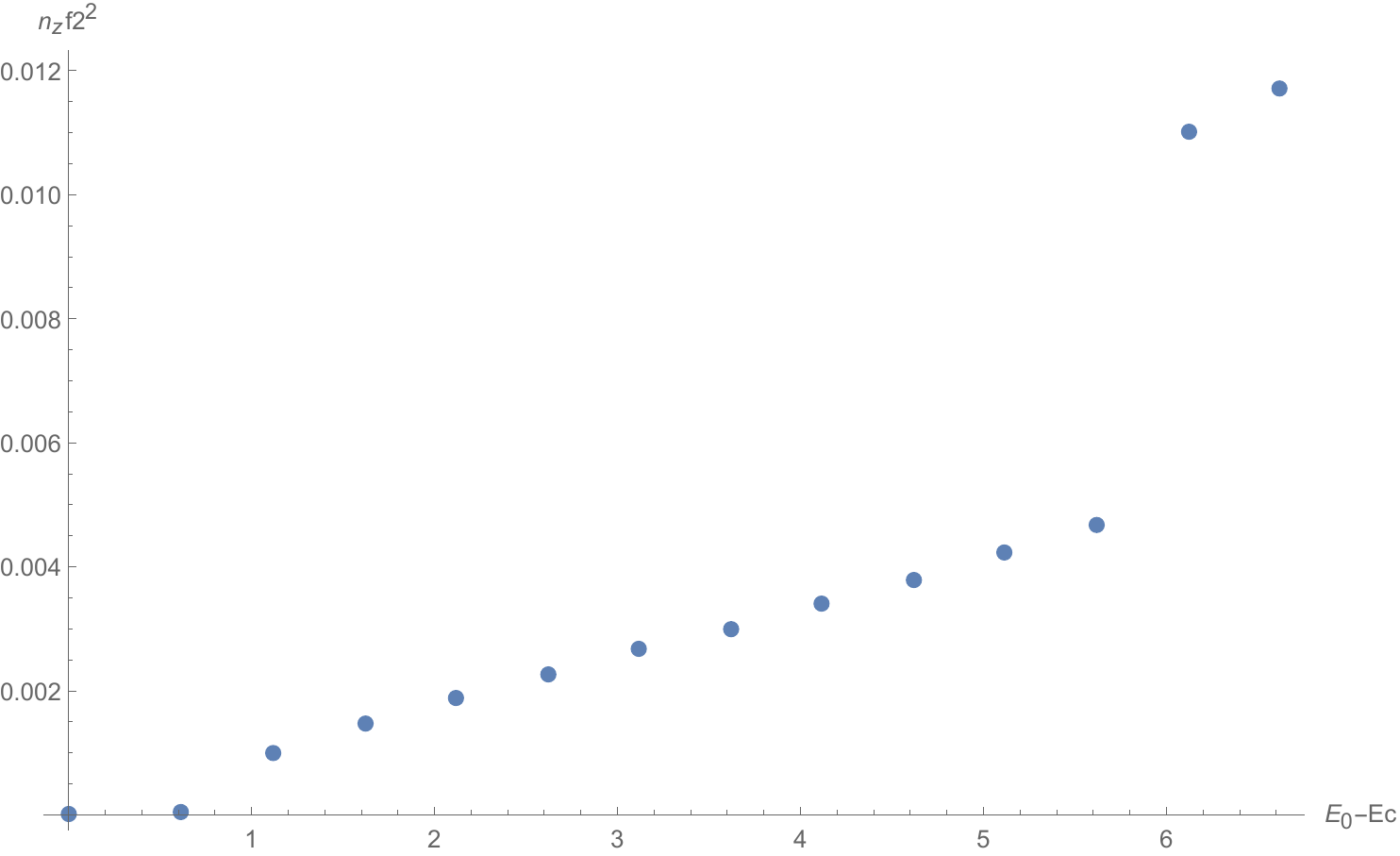}
\caption{The dependence of the squared amplitude of the main Fourier harmonic of $n_z$ on electric field for parameters presented in column C of the Table \ref{tab:tab1}. If field $E_{0c}<E_0<E_2$ excitation has wave vector about $q_2$, for larger electric field the wave vector of excitation is about $q_1$.}
\label{fig:re1}
\end{figure}

For parameters, presented in the column C of the Table \ref{tab:tab1}, the factor $\Lambda$ has two maxima with the positions $q_1\approx (2.2,4.9)~\mu m^{-1}$ and $q_2\approx (3.7,8.3)~\mu m^{-1}$. The value $\Lambda(q_2)$ is larger than $\Lambda( q_1)$. Above the threshold perturbations with wave vector $\bm q_2$ are increasing and then the stationary pattern appears. The squared amplitude of corresponding Fourier harmonic of $n_z$ as a function of field is presented in Fig. \ref{fig:re1}. As it is seen the transition at $E_{0c}$ is a continuous phase transition. If the electric field amplitude is less than some value $E_2$, the appearing structure has space period close to $2\pi/q_2$.
At $E_0 > E_2$ a second transition occurs and the period of the pattern becomes close to $2\pi/q_1$. This transformation is analogous to the first order transition. At these fields the modulation period of the pattern becomes  $\sim 1.7$ times larger. Note, that the used calculation method permits to study the phase behavior, but does not allow to determine the second transition point exactly, so the value of $E_2$ is an estimation.

Thus in the second region of the phase diagram at some parameters the system has the second order transition to the oblique stationary stripes. At larger electric field values the first order transition with the change of the pattern modulation period happens. At different parameters set a single weak first order transition to the stripes with the period close to the $2\pi/q_1$ takes place.

Now, let us examine solutions in the region III of the phase diagram presented in Fig. \ref{fig:pd}. In this region, the structures that appear have wave vectors with a non-zero $q_{cx} \ne 0$ and a complex increment. As shown in Ref. \cite{PK24}, the complexity of the increment is due to the hybridization of the electric potential mode with one mode of the director. The two modes that result from this hybridization have complex conjugated increments and give rise to the resulting excitation. There are two possible scenarios for the extrema: the absolute values of the amplitudes of the modes with conjugate increments are either equal or one of those amplitudes is zero. In the first case, we have a standing wave excitation, while in the second scenario, the resulting structure is a traveling wave.

\begin{figure}
\includegraphics[scale=0.4]{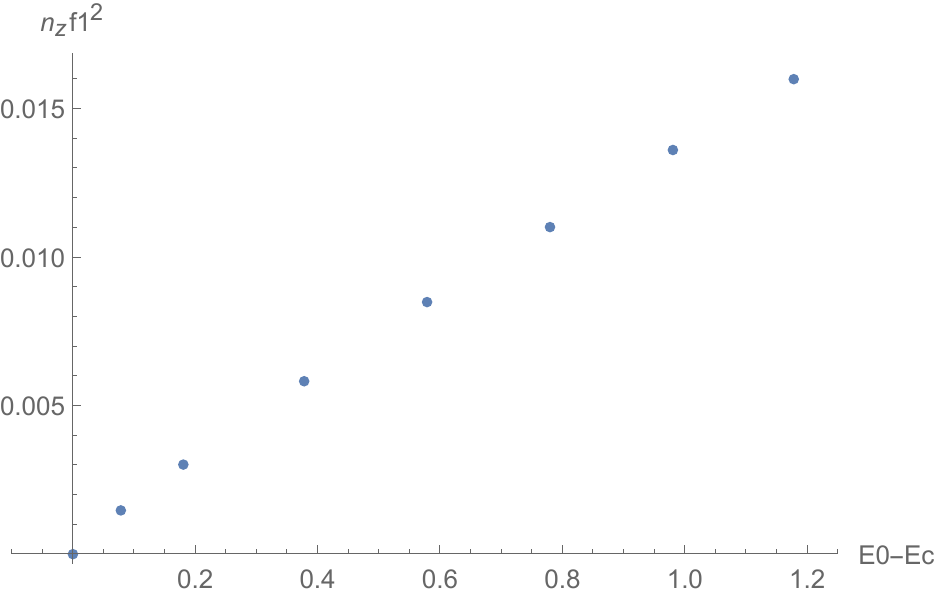}
\caption{The dependence of the squared amplitude of the main Fourier harmonic of $n_z$ on the electric field for the excitation with complex increment
in the standing wave. Parameters are presented in Table \ref{tab:tab1}, column D.}
\label{fig:stand}
\end{figure}

\begin{figure}
\includegraphics[scale=0.5]{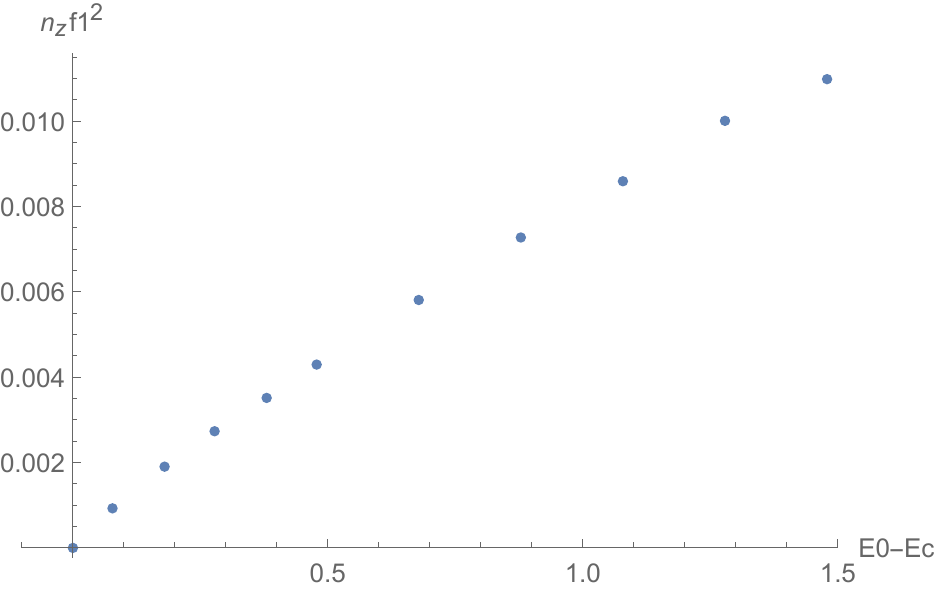}
\caption{The dependence of the squared amplitude of the main Fourier harmonic of $n_z$ on the electric field for the excitation with complex increment
in the traveling wave. Parameters are presented in Table \ref{tab:tab1}, column E.}
\label{fig:trav}
\end{figure}

We examined in detail the points D and E, see Fig. \ref{fig:pd}, the corresponding sets of parameters are provided in Table \ref{tab:tab1}, columns D and E. The point D corresponds to a standing wave. The dependence of $n_z$ on the amplitude of the electric field is presented in Figure \ref{fig:stand}. This pattern oscillates in time with the corresponding Hopf frequency. The point E corresponds to a traveling wave, the dependence of $n_z$ on the electric field is presented in Fig. \ref{fig:trav}. The observed phase velocity of the wave is about $3.6\cdot 10^{-5}~m/s$, what agrees well with the calculated from the linear analysis phase velocity $3.56\cdot 10^{-5}~m/s$. As it follows from the Figs. \ref{fig:pd}, \ref{fig:trav}, in both cases the bifurcation is soft, i.e. the amplitude of $n_z$ is proportional to $\sqrt{E_0-E_{0c}}$.
Let stress that  traveling waves propagate along the critical
wave vector $\bm{q}_c=(q_{c x}, q_{c y})$, i.e. in $\xi $ direction, that is perpendicular to the axis of the oblique linear patterns.

\begin{figure}
\includegraphics[scale=0.4]{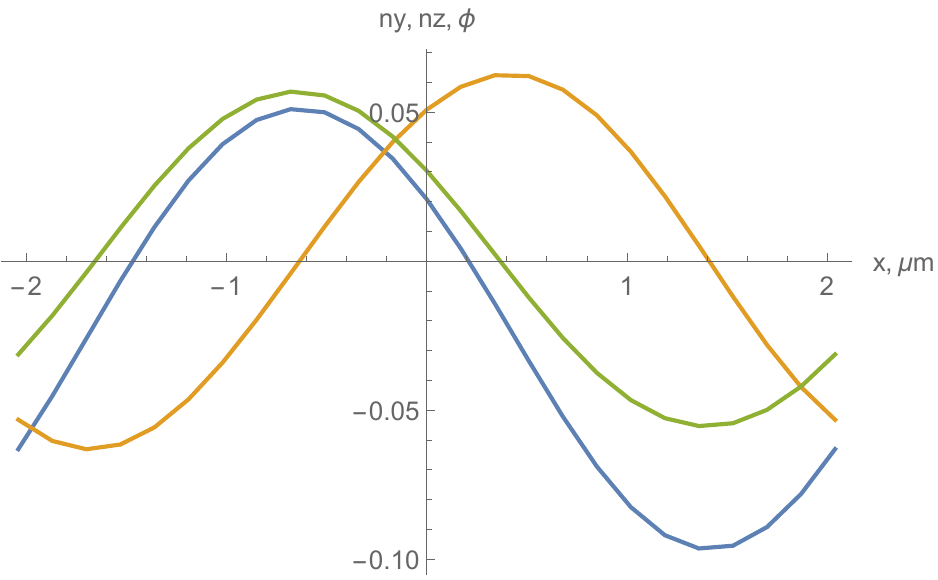}
\caption{The dependencies of $n_y$ (blue line), $n_z$ (yellow line) and potential $\Phi$
(green line, arbitrary scale) on the space coordinate in the traveling wave. Space period is about $4~\mu m$.}
\label{fig:nphi}
\end{figure}

\begin{figure}
\includegraphics[scale=0.3]{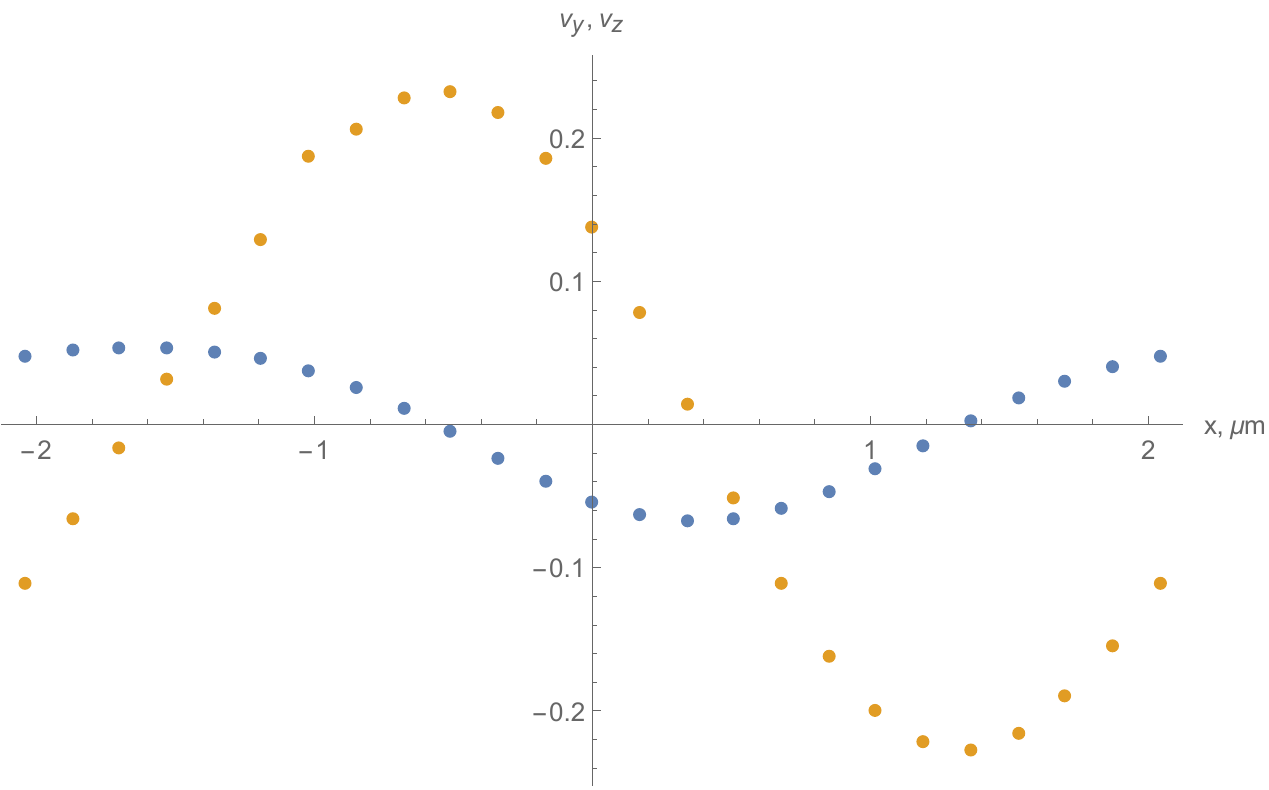}
\caption{The dependencies of $v_y$ (blue points) and $v_z$ (yellow points) in $cm/s$ on the space coordinate in traveling wave.}
\label{fig:vyvz}
\end{figure}

In a traveling wave, dynamic variables depend only on a linear combination of the space coordinate $\xi$ and time. For example, we present the dependencies of the variables on $x$ in a traveling wave with an electric field amplitude of $E_0-E_{0c} \approx 0.3$. The obtained dependencies of the director components $n_y$, $n_z$, and potential $\Phi$ on $x$ are shown in Fig. \ref{fig:nphi}. The blue line represents $n_y$, the yellow line represents $n_z$, and the green line represents $\Phi$. All three have an arbitrary scale. The figure shows the dependencies within one period, which is approximately 4 $\mu$m. It is easy to see that the average value of $n_y$ is nonzero. Analogous dependencies of the velocity components $v_y$ and $v_z$ on $x$ are shown in Fig. \ref{fig:vyvz}. The blue line represents $v_y$, and the yellow line represents $v_z$. The velocities are calculated in $cm/s$.

\begin{figure}
\includegraphics[height=3 cm]{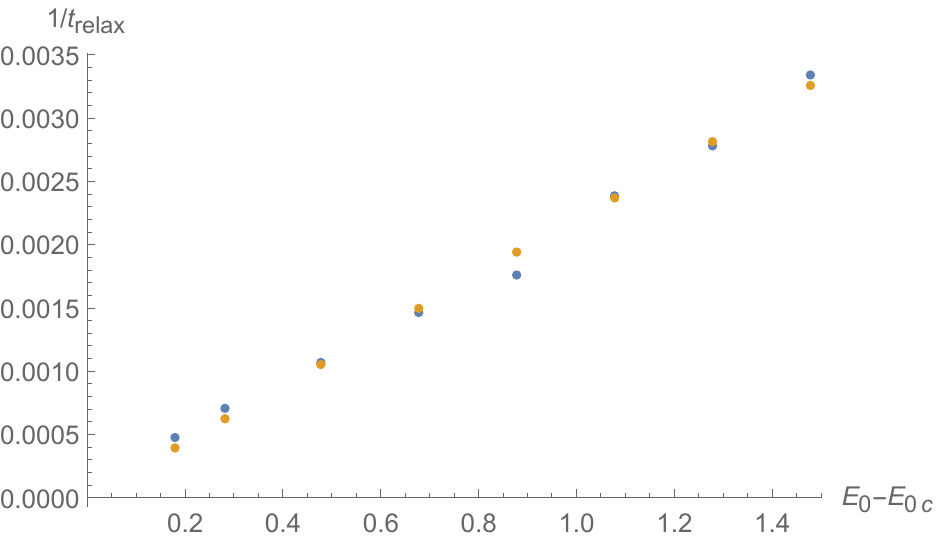}
\caption{Blue points present the dependence of  relaxation time, observed in calculation.
Yellow points are the linear function $\sim (E_0-E_{0c})$, fitted to the blue points.}
\label{fig:trelax}
\end{figure}

Figure \ref{fig:trelax} presents the dependence of inverse relaxation time of the system on the external field
for traveling wave. The time value was determined by the calculation of the relaxation started from small
random initial condition for all unknown fields. At the beginning of the process the appearing solution
has form of the standing wave, then during the subsequent relaxation it is transformed to the traveling wave.
The calculated dependence of Fourier harmonic amplitude of $n_z^2(t)$ on time was plotted as $\varphi(t)=\log (|n_z^2(t)/n_z^2(\infty )-1|)$. The relaxation time was determined from the slope of envelop of this dependence $\varphi(t)\sim const -t/t_{relax}$, its value is expressed in units of the external field period. In the figure blue points represent the inverse relaxation time and yellow points are the linear function $\sim (E_0-E_{0c})$ fitted to the obtained dependence.

For the convenience of readers, let's summarize the results of this section. In the region III of the diagram where the ratio $\eta/\gamma$ is small enough, the excitation, appearing above the instability threshold, has complex amplification factor and increment. In this case the critical mode is represented by two degenerate modes,
traveling in the opposite directions. Corresponding stable solution can be either a traveling or a standing wave.
The calculated diagram demonstrates that traveling mode exists in the larger part of the region III.
The stability region of standing mode is confined in both directions over parameters $\eta/\gamma$ and $\zeta/(K \epsilon_0 |\Delta\epsilon|)^{1/2}$. This pattern appears only at moderate values of the ratio $\eta/\gamma$ in the transient region between the stability region of the traveling mode and regions I and II.

The frequency of the external field influences the instability, in particular it changes wavevectors of the instability bifurcation. We illustrated this observation performing the one-dimensional computations for the points D and E in the region III. We found that the stable pattern at the frequency $400~Hz$ corresponds to the standing  wave in the point D and the traveling wave in the point E. At the decrease of the frequency the pattern in the point D remains the same, whereas for the point E at the much smaller frequency, $50\, Hz$, (but still within the assumptions of the minimal model $\omega \gg \sigma_\perp /\epsilon_\perp$ ) the stationary state is transformed to the standing wave. When the frequency is increased the pattern in the point E is qualitatively the same, but the pattern in the point D at the frequency $600~Hz$ is transformed to the traveling wave. Thus the increase of the external field frequency reduces the stability region of the standing wave patterns. However location of the boundary line (at our two-dimensional cross-section of the phase diagram, see figure \ref{fig:pd}) between the regions I and III only weakly depends on the frequency.

In the region II the increase of the external field frequency results in the shift of the boundary between the subregions of the first and second order transitions to the larger values of $\eta/\gamma$.

\section{Stability of the one-dimensional patterns with respect to two-dimensional perturbations}
\label{sec:two-dim}

As it was demonstrated in Sec. \ref{sec:nonlinear}, our problem is reduced to the set of local dynamic nonlinear equations (\ref{trun4},\ref{trun3},\ref{ro},\ref{potdyn2},\ref{trunp1}). Then, based on the inequality $q_cd \gg 1$, we neglect a $z$-dependence of all our fields.  Therefore, we stay with the set of two-dimensional equations. The equations allow us to verify the results obtained from the one-dimensional simulations. In particular, we are interested in the stability of the one-dimensional solutions that we have found.

 We study nonlinear dynamics of nematics slightly above the flexoelectric instability threshold $E_{0 c}$, i.e. under the condition $(E_0-E_{0 c})/E_{0 c} \ll 1 $ in frame of   the system of two-dimensional nonlinear equations (\ref{trun4},\ref{trun3},\ref{ro},\ref{potdyn2},\ref{trunp1}). In this case, the characteristic relaxation time to the stationary state is about $10^3$ to $10^4$ times larger than the period $T=2\pi/\omega$ of the external alternating electric field. For the two-dimensional simulations the introduced in Subsec. \ref{subsec:fictitious} parameters are chosen as follows:  $A^2 = 2.5  \cdot 10^{10}\,$ V$^2$ m$^{-2}$, $G_1 = \, 1.54 \cdot 10^{11} \, m^{-2}$. The parameters satisfy the necessary inequalities, see Subsec. \ref{subsec:fictitious}. The computational box is chosen to be $\{(-\pi/q_{cx}, \pi/q_{cx}),(-\pi/q_{cy}, \pi/q_{cy})\}$, where $q_{cx},q_{cy}$ are the components of the critical wave vector,  taken for a given set of material parameters. It can be found in the framework in the linear approximation and $q_c \sim (\gamma \omega /K)^{1/2}$, see Ref. \cite{PK24}. Let us remember that the tilt of the oblique patterns (i.e. the ratio of values $q_{c y}/q_{c x}$) increases as the control parameter $\eta/\gamma$ grows.

Thus we obtain that the evolutionary behavior of the nonlinear dynamical system is described by its relaxation to the equilibrium state during finite time with a critical dependency on the deviation from the instability threshold.
We arrive to the conclusion that above the bifurcation threshold any initial perturbations of the homogeneous state transform rather fast into a state dominated by the critical mode. The subsequent evolution is slow, it is explained by soft or near soft bifurcation. For the real instability increments the resulting patterns are static and one-dimensional.  We have verified over numerical calculations of  the  set of two-dimensional nonlinear equations, that  the static effective one-dimensional solutions,   obtained in previous section for the regions I and II of  the "phase diagram", see Fig. \ref{fig:pd},  are stable.  However, we  also find, that  the obtained solution in the form of an effective one-dimensional traveling wave loses stability  with respect to small two-dimensional perturbations and ceases to be effective one-dimensional. On the contrary, obtained  solution in the form of an effective one-dimensional standing wave stay stable under small two-dimensional perturbations.
This situation is approved by the phenomenological theory of  2D-case, see the end of Sec. \ref{sec:phen}. Thus we have verified that this type of solutions is stable under small perturbations in some region above the bifurcation threshold.
If we substitute the obtained solutions as the initial conditions for numerical calculations  below the instability threshold, then the solution  dies out in the agreement with the linear analysis.
Let  note also, that the regimes of the oblique rolls are characterized by the appearance of small nonzero average value  of the nematic director component $n_y$. This fact means the rotation of the undistorted nematic director in comparing with its initial state.

It is worth to pay some attention to the phase diagram region
near the transition between traveling and standing wave bifurcations (in terms of the Ginzburg-Landau phenomenology (see Sec. \ref{sec:phen}) the region corresponds to close values of the phenomenological coefficients $b_1^\prime $ and $b_2^\prime$). In such a region one should expect very slowly damped
crossover-like quasi-beats.   The quasi-beats are the result of  presence in the solution of a residual wave of small amplitude,  propagating in the opposite direction to the main one. The amplitude of the quasi-beats is smaller than the amplitude of the principal solution and decreases with distance from the discussed region in the phase diagram, for example, with decrease of  $\eta$.  Apparently it is the case for some more or less typical for liquid-crystalline material parameters. Based on this simple observation it is tempting to speculate on why propagating patterns observed experimentally in the  electro-hydrodynamic instabilities of $(- -)$ nematic liquid crystals typically have more complicated structure than
simple propagating rolls (see, e.g. \cite{ES16}, \cite{KD15}, \cite{TC07}).

\begin{figure*}
\hskip-0.1true cm	\includegraphics[scale=0.34]{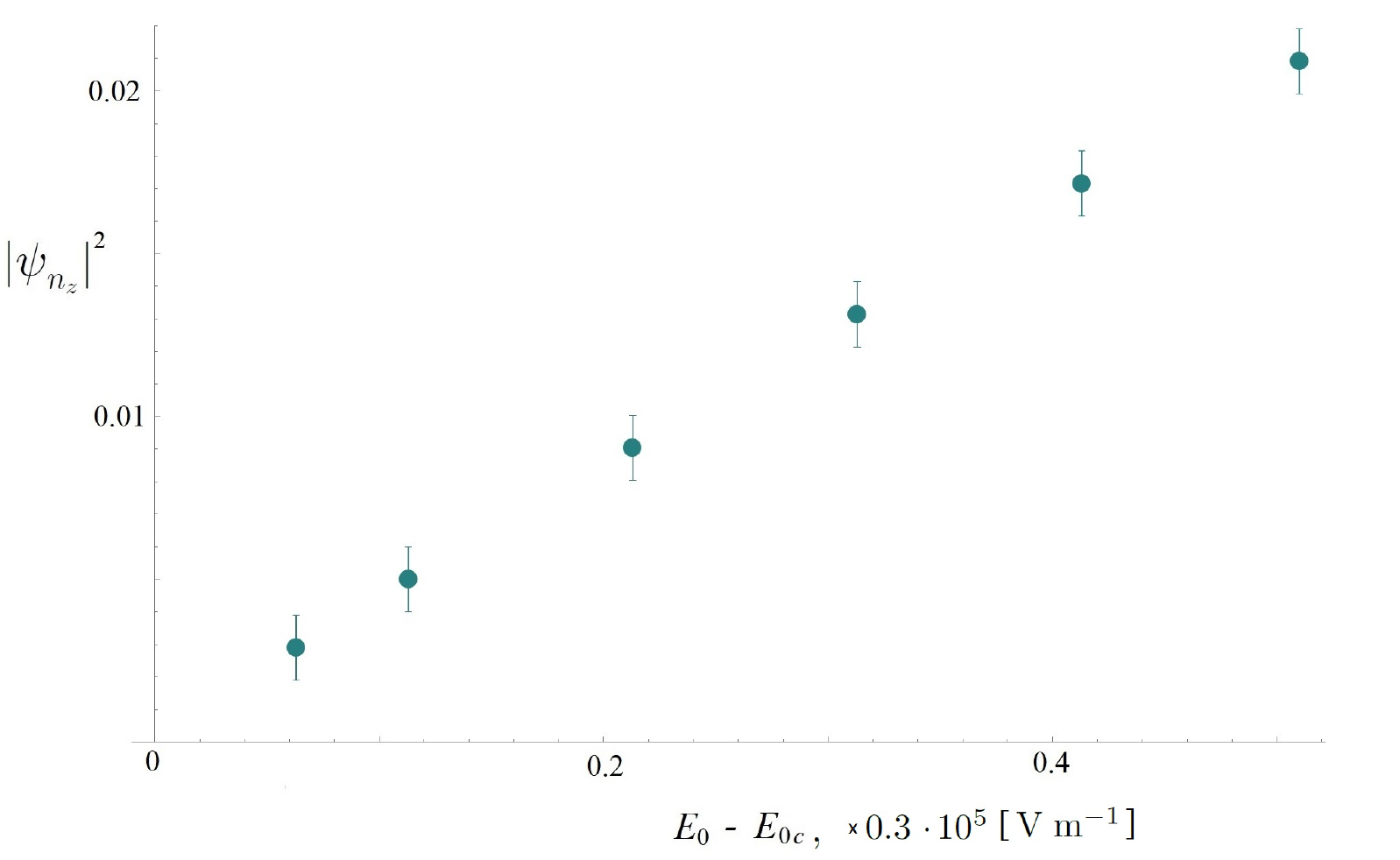}
\hskip0.1true cm
\includegraphics[width=0.45\columnwidth]{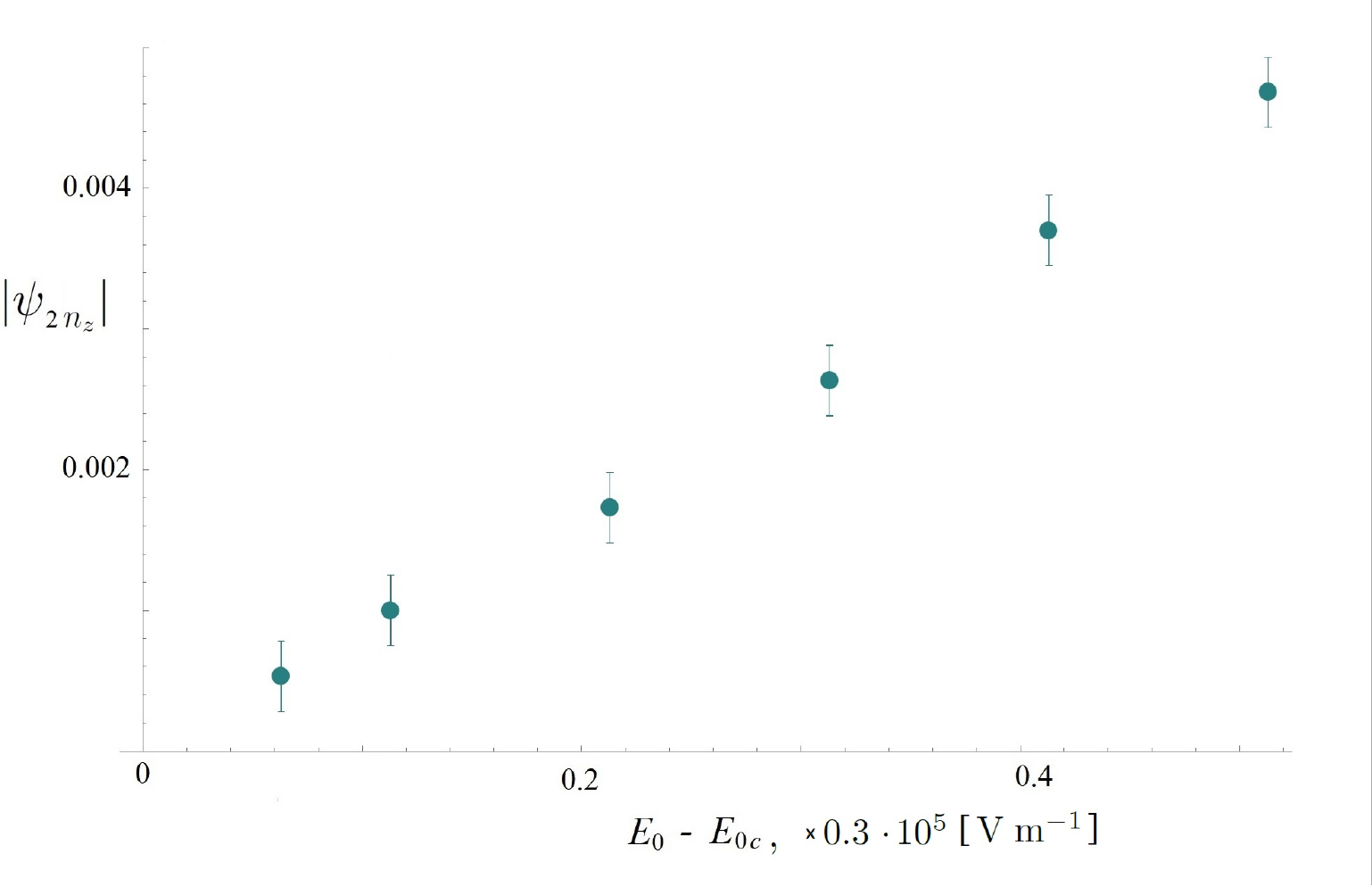}
\caption{ Dependencies  of the square of the first  harmonic   amplitude $|\psi_{n_z}|^2$ and absolute value of second  harmonic amplitude $|\psi_{2 n_z}|$ of the  solution of standing wave for $n_z $ over $(E_0-E_{0 c})$, where  $E_{0 c}= 36.116 \cdot 10^{5}\,$ V m$^{-1}$  for the  set of  material parameters:
		$ \omega $ (frequency) $ = 2\pi \cdot\,300\,$ s$^{-1}$;
 		$\sigma_\perp = 35 \cdot 10^{-10}$ $\Omega^{-1} \cdot m^{-1}$, $\Delta \sigma =  - 0.5 \sigma_\perp $,
		$K_1 = K_2 = K_3 = K = 6.0 \cdot 10^{-12}\,$ N,
		$\zeta = 6.75\cdot  10^{-11}$ ${C}\cdot m^{-1}$,
		${\epsilon}_\perp = 16 $,
		$\Delta {\epsilon} = - 4\,\pi \cdot 10^{-1}$,
		$\gamma = \, 0.06\, Pa \cdot s$, $\eta = 0.04\,Pa \cdot s$, $\mathcal{C} = 257.027 \cdot 10^4\, s \cdot m^{-2}$, $\mathcal{C}_p = 102.808 \cdot 10^4\, s \cdot m^{-2}$, $G_p = \, 6.17 \cdot 10^{10} \, m^{-2}$, $G_v =2.467  \cdot 10^{10} Pa\cdot s\cdot m^{-2}$, $q_{c x} = 10.205 \cdot  10^{5}\, m^{-1}$, $q_{c y} = 28.26 \cdot  10^{5}\, m^{-1}$.   }
\label{trcr}
\end{figure*}

Performing 2D-numerical calculations we also verify the critical character of the amplitudes of the solutions  first and second harmonics dependencies on the deviation of the external electric field
from the threshold point. Thus, the results of the computations for  point from region III of the "phase diagram", are presented in Fig. \ref{fig:pd}.
The  dependencies of the square of  first  harmonic   amplitude $|\psi_{n_z}|^2$ and absolute value of second  harmonic amplitude $|\psi_{2 n_z}|$ of $n_z $ for the regime  of oblique patterns standing wave over $(E_0-E_{0 c})$ are presented  in Fig. \ref{trcr} for the cases of single-Frank-constant approximation    for certain set of other material parameters. For the case of different $K_\alpha$ we have obtained the similar  results.  The equations for the case of different Frank modules are described in Appendix \ref{sec:A}.

Note, that the greater the value of the imaginary part of the characteristic exponent, the greater is the Hopf oscillation frequency (or speed if the texture is propagating one)  of time-dependent patterns. The most noticeable increase in the imaginary part of the characteristic exponent is achieved with an increase in electrical conductivity, obviously if the corresponding condition of applicability is satisfied,  and with a decrease of the control parameter $\eta/\gamma$.
Remember also, that we investigated numerically the dependencies of the critical electric field $E_{0 c}$ and of the critical wave vector on the frequency $\omega$ in \cite{PK24}, providing the corresponding typical dependencies. We confirmed that the critical wave vector and the critical electric field amplitude depend as $\sqrt{\omega}$ on the frequency. Thus the value of the critical vector increases together with the critical field amplitude.

In addition, let us check that the obtained results are independent of the values of the additional parameters involved in  Sec. \ref{sec:nonlinear}, \ref{subsec:fictitious}, if they satisfy the conditions, given in the text.  Let stress that nothing at all depends on the coefficients at the time derivatives of potential and pressure,  that are added manually to the right side of the equations (\ref{potdyn}), (\ref{trunp1}),  correspondingly, if they satisfy the  smallness  conditions given in the text, see Sec. \ref{subsec:fictitious}.
The contributions  proportional to coefficient $A$, that we add to equations to suppress the zero mode,
must satisfy the condition of smallness in comparison to those that are proportional to the  second derivatives in the $x$- and $y$-directions, i.e. $\partial^2_x$ and $\partial^2_y$, respectively.
Generally speaking, these contributions somewhat weakly shift the instability threshold. This effect is expected, since the introduction of these contributions replaces the boundary conditions along $z$-axis. If the discussed smallness condition of the terms is satisfied, then the shift is small, and discussed contributions  have no effect on anything other than zero-mode suppression.

\section{Phenomenological theory of pattern formations}
\label{sec:phen}

The complete set of non-linear electro-hydrodynamic equations in nematics is too complicated to be solved analytically or even numerically. It contains a large number of material parameters purely determined experimentally. Fortunately, to  gain insight into possible types of bifurcations and patterns formed, a Ginzburg-Landau-like phenomenology can be applied. In such approach, valid for a weakly non-linear regime, the pattern formation is determined by a few phenomenological parameters, which in principal can be determined by confronting phenomenological theory predictions with the results of numeric simulations or experimental data.

Then, the dynamics of the critical mode can be effectively separated from other degrees of freedom, and therefore can be analyzed using standard bifurcation techniques. We consider the dynamics of nematics above the flexoelectric instability threshold on times much larger than the period $2\pi/\omega$ of the external alternating electric field. The approach is justified by long relaxation times of the critical mode near the instability point.

The principal peculiarity of the phenomenology is that the critical mode has a finite wave vector $\bm q$. The property is closely related to the character of the flexoelectric instability. Let us designate the amplitude of the critical mode as a complex parameter $\psi$. One can assume that $\psi$ determines the main contribution to the $z$-component of the director:
\begin{equation}
n_z=\mathrm{Re}\, \left[
\psi \exp(i \bm q \bm r)\right] + \dots ,
\label{phen1}
\end{equation}
where dots designate subleading terms. The goal of the phenomenology is to establish a closed dynamic equation for the complex quantity $\psi$ and to find stable patterns on times larger than the period of the external electric field.

For the regions I and II in the diagram drawn in Fig. \ref{fig:pd} the final pattern is a stationary one-dimensional pattern. The situation can be described by the following nonlinear equation for the amplitude $\psi$
\begin{equation}
\partial_t\psi= \lambda \psi - b |\psi|^2 \psi.
\label{phen2}
\end{equation}
The structure of the equation can be explained by the selection rules that are dictated by the projection of the complete system of nematodynamic equations onto $\psi$, and by the perturbation expansion that is justified by the proximity to the instability threshold. The main rule is the conservation of the total wave vector. The equation (\ref{phen2}) implies that the real part of $b$, $b'$, is positive. Otherwise there are no stable solutions of the equation. (Here and further we denote the real part of some variable $b$ as $b'\equiv \mathrm{Re} (b)$.) The solution is zero if $\lambda'<0$ and non-zero if $\lambda'>0$ where $\lambda'$ is the real part of $\lambda$.

One could take analyze the mode $\phi$ with the double wave vector $2\bm q$. It determines the following contribution to $n_z$:
\begin{equation}
\mathrm{Re}\, \left[
\phi \exp(2i \bm q \bm r)\right],
\nonumber
\end{equation}
like in Eq. (\ref{phen2}). One can formulate the joint system of equations for $\psi,\phi$:
\begin{eqnarray}
\partial_t\psi= \lambda \psi - b_1 |\psi|^2 \psi -h_1 \phi \psi^\star \, ,
\label{phen3} \\
\partial_t \phi =\lambda_2 \phi - g_2 \psi^2 \, ,
\label{phen4}
\end{eqnarray}
with $\lambda_2 < 0$. Since the dynamics of $\phi$ doesn't undergo the critical slowing, near the instability point one can use the adiabaticity condition $\phi =(g_2/\lambda_2)\psi^2$, following from Eq. (\ref{phen4}) if the time derivative is neglected. Substituting the expression for $\phi$ into Eq. (\ref{phen3}), we return to Eq. (\ref{phen2}) with
\begin{equation}
b=b_1+h_1g_2/\lambda_2.
\label{phon5}
\end{equation}
Note that $\phi$ appears to be quadratic in $\psi$. The same logic can be applied to the zeroth harmonic which is also quadratic in $\psi$.

The coefficient $\lambda_2$ in the equation (\ref{phen4}) could be small. Therefore the interaction of the first and second modes can lead to the essential renormalization of the constant $b$ in Eq. (\ref{phen2}) in accordance with Eq. (\ref{phon5}). For some values of the parameters the $b'$ can become negative. That leads to a rigid bifurcation. The situation seems to be realized in the region II of the phase diagram depicted in Fig. \ref{fig:pd}. For several points in the region, which we have examined, hard bifurcations were observed.

To examine the rigid bifurcation where the real part of the cubic constant $b$ in Eq. (\ref{phen2}) is negative one should add to the equation the fifth-order term. This procedure is correct provided $b$ is sufficiently small. With the additional term taken into account, one arrive at the equation
 \begin{equation}
\partial_t\psi= \lambda \psi - b |\psi|^2 \psi - \gamma |\psi|^4 \psi\, .
\label{phen2-3}
\end{equation}
The fifth order term provides stabilization of $\psi$ if $\gamma'>0$, where $\gamma'$ is the real part of $\gamma$. The equation (\ref{phen2-3}) leads to the following equation for the absolute value of $\psi$:
\begin{equation}
\partial_t|\psi|= \lambda' |\psi| - b' |\psi|^3 - \gamma' |\psi|^5 .
\label{phen2-4}
\end{equation}
Equating to zero the right hand side of Eq. (\ref{phen2-4}) we find the stationary solution
\begin{equation}
|\psi|^2=\frac{1}{2\gamma'}
\left(-b'+\sqrt{|b'|^2+4\gamma' \lambda'}\right).
\label{phen2-5}
\end{equation}
The solution with the sign minus at the square root existing at negative $\lambda'$ is unstable and we don't consider it. The solution (\ref{phen2-5}) exists provided $|b'|^2+4\gamma' \lambda'>0$, in particular, it exists at negative $\lambda'$ if $4\gamma' |\lambda'|>|b'|^2$. Thus in the region two stable solutions exist and a hysteresis should be observed.

The analogous analysis should be done if $b'$ is positive, but is anomalously small. The bifurcation point where the instability changes from a soft (continuous) to the hard (jump-like) is an isolated point.
The case corresponds to an effective tricritical behavior, in terms of the phase transition theory. Then the nonzero stationary solutions are absent if $\lambda'<0$ and the expression (\ref{phen2-5}) determines the stable solution if $\lambda'>0$. Note that the expression (\ref{phen2-5}) is zero if $\lambda'=0$ that is in the instability point. Therefore, we are dealing with a soft bifurcation and the hysteresis is absent in this case. Such situation could be realized in the right part of the region II of the phase diagram depicted in Fig. \ref{fig:pd}. However, the possibility is screened by the hard bifurcation for the competing mode with a wave vector that differs from the wave vector of the basic critical mode, see Sec. \ref{sec:one-dim}.

The discussion presented above is applied to the case of a purely decaying dynamics of the critical mode realized in the regions I and II  of the phase diagram depicted in Fig. \ref{fig:pd}. The situation with oscillating dynamics realized in the region III is slightly more complex. Then we should generalize the definition (\ref{phen1}) to obtain
\begin{equation}
n_z=\mathrm{Re}\, \left\{
[\psi_+(t) +\psi_-(t)]\exp(i \bm q \bm r)\right\},
\label{runs1}
\end{equation}
where the complex parameters $\psi_+$ and $\psi_-$ contain positive and negative frequencies, respectively. The selection rules are even stronger in this case since one should take into account the frequency conservation law besides the wave vector conservation law. However, the general scheme remains the same, leading to a universal dynamic equations for the amplitudes $\psi_{\pm}$ of the critical mode.

In the main approximation, the dynamic equations for $\psi_{\pm}$ are written as
\begin{eqnarray}
\partial_t \psi_+=\lambda \psi_+ -b_1 |\psi_+|^2 \psi_+
-b_2 |\psi_-|^2 \psi_+,
\label{runs2}  \\
\partial_t \psi_-=\lambda^\star \psi_- -b_1^\star |\psi_-|^2 \psi_-
-b_2^\star |\psi_+|^2 \psi_-,
\label{runs3}
\end{eqnarray}
where $\lambda, b_1, b_2$ are complex factors. We assume that $\lambda''=\mathrm{Im}\, \lambda <0$, the condition corresponds to positive frequencies in $\psi_+$. The equations (\ref{runs2},\ref{runs3}) can produce both traveling waves and standing waves.

At large times the equations (\ref{runs2},\ref{runs3}) lead to an asymptotic state which has an oscillating in time behavior of $\psi_+$ and $\psi_-$. The conditions for the asymptotic state following from the Eqs. (\ref{runs2},\ref{runs3}) have form
\begin{eqnarray}
\lambda' - b_1' |\psi_+|^2 -b_2'|\psi_-|^2=0,
\label{runs4}  \\
\lambda' - b_1' |\psi_-|^2 -b_2'|\psi_+|^2=0,
\label{runs5}
\end{eqnarray}
where it is assumed that $\lambda', b_1', b_2'$ are positive. Obviously, the system of equations (\ref{runs4},\ref{runs5}) has a solution
\begin{eqnarray}
|\psi_+|=|\psi_-|= \sqrt\frac{\lambda'}{b_1'+b_2'},
\label{runs6}
\end{eqnarray}
corresponding to the standing wave. Besides, there is a solution of Eqs. (\ref{runs2},\ref{runs3})
\begin{eqnarray}
|\psi_+|=\sqrt{\lambda'/b_1'}, \quad
\psi_-=0,
\label{runs7}
\end{eqnarray}
corresponding to the traveling wave.

The equations for the absolute values of $\psi_+$ and $\psi_-$ are separated from the equations for the phases, they have form
\begin{eqnarray}
\partial_t |\psi_+|=\lambda' |\psi_+| -b_1' |\psi_+|^3
-b_2' |\psi_-|^2 |\psi_+|,
\label{runs8}  \\
\partial_t |\psi_-|=\lambda' |\psi_-| -b_1' |\psi_-|^3
-b_2' |\psi_+|^2 |\psi_-|.
\label{runs9}
\end{eqnarray}
The stability analysis of the equations (\ref{runs8},\ref{runs9}) shows that if $b_1'>b_2'$, the solution (\ref{runs6}) is stable and the solution (\ref{runs7}) is unstable. In the opposite case $b_2'>b_1'$ the solution (\ref{runs7}) is stable and the solution (\ref{runs6}) is unstable. Near the stationary state the time dependence of small perturbations of $|\psi_+|$ and $|\psi_-|$ have form
$\propto \exp(-\alpha t)$. The decrements $\alpha$ which determine this relaxation are
\begin{equation}
\alpha_1=2\lambda', \quad
\alpha_2=2\frac{b_1'-b_2'}{b_1'+b_2'} \lambda',
\label{runs11}
\end{equation}
for the standing wave and
\begin{eqnarray}
\alpha_1=2\lambda', \quad
\alpha_2=\frac{b_2'- b_1'}{b_1'}\lambda',
\label{runs10}
\end{eqnarray}
for the traveling wave. The decrements are of the order of $\lambda'$.

The critical behavior of $\lambda'$ near the threshold $E_c$ is determined by the law $\lambda'\propto E-E_c$. Therefore the amplitudes of $\psi_{\pm}$ (\ref{runs6},\ref{runs7}) are proportional to $\sqrt{E-E_c}$. The instability increments are proportional to $E-E_c$. The frequency of the waves is determined mainly by the imaginary part of $\lambda$, since the corrections to the frequency related to the imaginary parts of $b_1, b_2$, are proportional to $\lambda'$ that is to $E-E_c$. Thus the frequency has no critical behavior.

Note that the decrements $\alpha_2$ in Eqs. (\ref{runs11},\ref{runs10}) tend to zero as $b_1' \to b_2'$, indicating the existence of a slow mode. Indeed, it follows from Eqs. (\ref{runs8},\ref{runs9})
\begin{eqnarray}
\partial_t \left(|\psi_+| /|\psi_-|\right)
=(b_1'-b_2')|\psi_+| |\psi_-|
\nonumber \\
+(b_2'-b_1') |\psi_+|^3/ |\psi_-|.
\nonumber
\end{eqnarray}
Thus, near the point where $b_1'=b_2'$ the ratio $|\psi_+| /|\psi_-|$ evolves slowly. This phenomenon can be called critical slowness, by analogy with second order phase transitions. The slowness has to be observed on top of the critical behavior near the threshold.

Next important step is to check whether the one-dimensional solutions (\ref{runs6},\ref{runs7}) are stable
with respect to transverse (two-dimensional) perturbations. To investigate the possibility we need
to develop two-dimensional phenomenology. Then the expressions (\ref{phen1},\ref{runs1}) should be further generalized to include Fourier harmonics with two equivalent critical wave vectors, $\bm q_1$ and $\bm q_2$, having equal $x$-components and opposite $y$-components. This case corresponds to the tilted wave vector of the critical mode and is realized, particularly, in the region III of the phase diagram depicted in Fig. \ref{fig:pd}. In this case the contribution to $n_z$ related to the critical mode can be expressed as
\begin{equation}
\mathrm{Re}\, \left\{
[\psi_{1 +}(t) +\psi_{1 -}(t)]\exp(i \bm q_1 \bm r)
+  [\psi_{2 +}(t) +\psi_{2 -}(t)]\exp(i \bm q_2 \bm r)\right\},
\label{runs111}
\end{equation}
where $\psi_{... +}$ and $\psi_{... -}$ contain positive and negative frequencies, respectively.

The equations for the functions $\psi_{1 +}(t), \psi_{1 -}(t)$, $\psi_{2 +}(t), \psi_{2 -}(t)$ can be derived by analogy with  deriving Eqs. (\ref{runs2},\ref{runs3}). Taking into account all the selection rules, we arrive at the following equations
\begin{eqnarray}
\partial_t \psi_{1 +}=\lambda \psi_{1 +} - b_1 |\psi_{1 +}|^2 \psi_{1 +}
- b_2 |\psi_{1 -}|^2 \psi_{1 +} - b_3 |\psi_{2 +}|^2 \psi_{1 +}
- b_4 |\psi_{2 -}|^2 \psi_{1 +} - b_5\,  \psi_{2 +} \psi^\star_{2 -} \psi_{1 -}  \, ,
\label{2d1}  \\
\partial_t \psi_{1 -}=\lambda^\star \psi_{1 -} -b_1^\star |\psi_{1 -}|^2 \psi_{1 -}
-b_2^\star |\psi_{1 +}|^2 \psi_{1 -} -   b_3^\star |\psi_{2 -}|^2 \psi_{1 -}
- b_4^\star |\psi_{2 +}|^2 \psi_{1 -}   - b_5^\star\,  \psi_{2 -} \psi^\star_{2 +} \psi_{1 +}   \,,
\label{2d2}\\
\partial_t \psi_{2+} = \lambda \psi_{2 +} - b_1 |\psi_{2 +}|^2 \psi_{2 +}
- b_2 |\psi_{2 -}|^2 \psi_{2 +} - b_3 |\psi_{1 +}|^2 \psi_{2 +}
- b_4 |\psi_{1 -}|^2 \psi_{2 +} - b_5\,  \psi_{1 +} \psi^\star_{1 -} \psi_{2 -} \,,
\label{2d3} \\
\partial_t \psi_{2-} = \lambda^\star \psi_{2 -} - b_1^\star |\psi_{2 -}|^2 \psi_{2 -}
- b_2^\star |\psi_{2 +}|^2 \psi_{2 -} - b_3^\star |\psi_{1 -}|^2 \psi_{2 -}
- b_4^\star |\psi_{1 +}|^2 \psi_{2 -} - b_5^\star\,  \psi_{1 -} \psi^\star_{1 +} \psi_{2 +}\, .
\label{2d4}
\end{eqnarray}
The equations are invariant under the permutation of the wave vectors $\bm q_1$ and $\bm q_2$.

Let consider the case of zero phase shifts between $\psi_{1 }$ and $\psi_{2 }$ functions.
We still assume that $\lambda'>0,b_1'>0,b_2'>0$. Then the system of equations (\ref{2d1}-\ref{2d4}) has asymptotic solutions
\begin{eqnarray}
|\psi_{1 +}|= |\psi_{1 -}| = \sqrt\frac{\lambda'}{b_1'+b_2'} , \quad
\psi_{2 +}=0 ,\,  \psi_{2 -}=0 \, ,
\label{run2d2}
\end{eqnarray}
and
\begin{eqnarray}
|\psi_{1 +}|=\sqrt{\lambda'/b_1'}, \quad
\psi_{1 -}=0 , \, \psi_{2 +}=0 ,\, \psi_{2 -}=0 \, ,
\label{run2d1}
\end{eqnarray}
corresponding to the one-dimensional standing and traveling wave, respectively. The solutions (\ref{run2d2},\ref{run2d1}) are the solutions (\ref{runs6},\ref{runs7}) written in terms of the single Fourier harmonic.

Stability analysis of the solutions (\ref{run2d2},\ref{run2d1}), based on the equations (\ref{2d1}-\ref{2d4}), shows that small perturbations on top of the traveling wave (\ref{run2d1}) have the following decrements
\begin{eqnarray}
\alpha_1=2\lambda' \, , \quad
\alpha_2= \frac{b_2'- b_1'}{b_1'}\lambda' \, , \quad
\alpha_3= \frac{b_3'- b_1'}{b_1'}\lambda' \, , \quad
\alpha_4=  \frac{b_4'- b_1'}{b_1'} \lambda' \, .
\label{runs2da}
\end{eqnarray}
The first two decrements coincide with the expressions (\ref{runs10}), and the decrements $\alpha_3,\alpha_4$ characterize two-dimensional perturbations. Small perturbations on top of the standing wave (\ref{run2d2}) have the following decrements
\begin{eqnarray}
\alpha_1, \quad
-2\, \frac{b_1'}{b_1'+b_2'} \alpha_2 ,
 \quad
\alpha_5=\Big(\frac{b_3'+b_4' - b_5' - (b_1'+ b_2') }{b_1'+ b_2'} \Big) \,\lambda' \, ,\quad
 \alpha_6= \Big(\frac{b_3'+b_4' + b_5' - (b_1'+ b_2') }{b_1'+ b_2'} \Big) \,\lambda' \, .
\label{runs2da}
\end{eqnarray}
 and the solution (\ref{run2d2}) is stable if $b_1'>b_2'$ and $min\{(b_3'+b_4'  - b_5'),(b_3'+b_4'  + b_5')\} >(b_1'+ b_2')$, thus two conditions are more likely to be fulfilled  simultaneously than three for stability of effective 1D-traveling wave. The situation described in Sec. \ref{sec:two-dim}, when the standing wave is stable under small two-dimensional perturbations whereas the traveling wave is unstable, corresponds to different signs of $\alpha_3,\alpha_4$ and to positiveness of $min\{\alpha_5,\alpha_6\}$.

\bigskip

\section{Conclusion}
\label{sec:conclusion}

In this work, presumably for the first time, we study numerically nonlinear stages of electro-hydrodynamic pattern formation in nematic liquid crystals under an external alternating electric field. We have limited ourselves to only one class of materials, which are $(- -)$ nematics. In this case there is the only physical mechanism, which is flexoelectricity, leading to instability of the homogeneous configuration of the nematic at the increase of the external electric field. The results reported in our paper are valid for relatively small above-threshold values of the electric field.

Being interested mainly in qualitative features of the pattern formation above the bifurcation threshold, we work within the framework of our previously proposed minimal model of electro-hydrodynamic instabilities in nematics. This model assumes a single-constant approximation for nematic elasticity and a two-constant approximation for viscous dissipation in the nematic, the single shear viscosity coefficient $\eta$ and director rotation friction coefficient $\gamma$. We do believe that a more realistic description will not affect our qualitative conclusions. We have checked this belief performing some calculations for the non-equal Frank coefficients. Guided by the prejudice that transparency is worth a few oversimplifications we restrict ourselves to the assumptions of the minimal model.

We examine the pattern formation above the flexoelectric instability threshold $E_{0 c}$ on time scales that are much larger than the period of the external electric field. Under these conditions, the emerging patterns can be classified using standard bifurcation theory, which was developed for systems that are homogeneous in time. To underlay the basic physical principles behind the pattern formation, our numerical results are rationalized and described by means of generalized Ginzburg-Landau phenomenological theory. The theory is valid near the bifurcation threshold and is independent of details of the instability. It enables one to describe all possible patterns in terms of a few phenomenological parameters entering the theory. Potentially, the phenomenological theory can be used to examine such non-linear objects as solitons.

We use the standard setup where the nematic film is placed between the capacitor plane plates and an alternative voltage is applied to the capacitor. We analyze the case where $q_c\, d >> 1$ where $q_c$ is the critical vector (the wave vector of the unstable mode) and $d$ is the film thickness (distance between the plates). In the limit the problem is reduced to a two-dimensional one, that is all fields can be treated as functions of the coordinates $x,y$ measured in the directions parallel to the capacitor plates. We solve numerically the set of non-linear electro-hydrodynamic equations with the periodic boundary conditions in the $x - y$ plane. The periodicity is selected to match the wave vector $\bm q_c$ of the critical mode, obtained in the linear approximation.

We have studied a range of more or less realistic material parameters and experimental conditions for which linear stability analysis \cite{PK24} predicts different types of instability. The "phase diagram" found in our numerics is drawn in Fig. \ref{fig:pd}, where different types of patterns are presented. We observe textures, appearing above the instability threshold, which are static one-dimensional patterns, standing (oscillating in time) patterns and traveling ones. The bifurcations that lead to the creation of textures can be either soft or rigid. Thus, we encounter a fairly complicated behavior. Let us describe the results in more detail.

The simplest behavior is observed in the region I. Here static one-dimensional  patterns, straight stripes, are realized above the instability threshold, where all the fields vary along the $x$-axis. The $x$-axis is chosen to be oriented along the director orientation below the instability threshold.  In the region II, where tilted static one-dimensional rolls are realized the situation is more complicated. We, rather unexpectedly, observed a type of a bifurcation that had apparently been overlooked and not previously discussed. This is a jump-like bifurcation with a change in the one-dimensional pattern period occurring in the part of the region II. In the other part of region II, a hard bifurcation occurs, accompanied by the condensation of the principal mode. In the regions I and II the resulting patterns are stable under perturbations including two-dimensional ones.

In the region III of the "phase diagram" the instability is described by a Hopf bifurcation. For the one-dimensional dynamic patterns we found the standard critical behavior of the principal spatial harmonic, its amplitude behaves as $\sqrt{E_{0}-E_{0c}}$ where $E_0$ is the amplitude of the alternating electric field and $E_{0c}$ is its critical value. One observes also the critical slowing, the relaxation time behaves as $(E_{0}-E_{0c})^{-1}$. Both laws are in agreement with the Ginzburg-Landau phenomenology. Near the boundary separating the regions of one-dimensional standing waves and travelling ones we find some extra slowness of the nematic dynamics. This phenomenon is also described within the phenomenology. We are confident that the results of our work are valid even beyond the minimal model. For the purpose we performed calculations taking elastic anisotropy, i.e. with three different Frank modules. The results are qualitatively the same.

In the frame of our numerical simulations we found that the traveling waves are unstable with respect to two-dimensional perturbations of the fields. On the contrary, the standing waves remain stable under small two-dimensional perturbations, they could be unstable under finite two-dimensional perturbations. The conclusions are based on the results of the simulations for a restricted set of parameters. It is not excluded that they need a correction for other sets. To analyze possible stable dynamic patterns, it is necessary to conduct a complete investigation of the two-dimensional solutions above the flexoelectric instability threshold. This work is currently in progress.

\acknowledgments

The work of E.S.P. and V.V.L. was supported by the Russian Science Foundation (Grant No. 23-72-30006),  and their work connected with the investigation of the case of different  Frank elastic modules  was supported  by the Ministry of Science and Higher Education of the Russian Federation (State assignment No. FFWR-2024-0014) of  Landau Institute for Theoretical Physics of the RAS.  The work of A.R.M. was supported by the Ministry of Science and Higher Education of the Russian Federation No. FMME-2025-0010 (No. 125020501404-4).

\appendix

\section{Details of calculations of the nonlinear equations}
\label{sec:A}

As we discuss in Sections \ref{sec:nonlinear}, \ref{sec:two-dim}, we  omit the terms proportional to $\partial_z$, and   Eq. (\ref{trun2}) can be written as:
\begin{eqnarray}
\partial_t \big(\partial_x D_x \big) + \partial_t \big(\partial_y D_y \big) = -\partial_k(v_k \nabla D)
- \partial_i(\sigma_{ik}E_k),
\label{DinD1}
\end{eqnarray}
where
\begin{eqnarray}
D_{x} =  \epsilon_0 \,(\epsilon_\parallel-\epsilon_\perp) (1 - n_y^2 - n_z^2)^{1/2} n_z E_0(t) - \epsilon_0 \, \epsilon_\parallel  \partial_x \Phi
+ \epsilon_0 \,(\epsilon_\parallel-\epsilon_\perp) ( n_y^2 + n_z^2) \partial_x \Phi
- \zeta \, ( n_y \partial_x n_y + n_z \partial_x n_z)
\nonumber \\
 -\,\epsilon_0 \, (\epsilon_\parallel-\epsilon_\perp) \,(1 - n_y^2 - n_z^2)^{1/2}\, (n_y \partial_y \Phi) + \zeta \, (1 - n_y^2 - n_z^2)^{1/2}\, ( \partial_y n_y)
\, , \
\nonumber \\
D_{y} = \, \epsilon_0 \,(\epsilon_\parallel-\epsilon_\perp) n_y  n_z E_0(t) - \epsilon_0 \, \epsilon_\perp \, \partial_y \Phi
- (\epsilon_\parallel-\epsilon_\perp)  n_y^2 \,\partial_y \Phi
\nonumber \\
 -\, \epsilon_0 \,(\epsilon_\parallel-\epsilon_\perp)\, n_y\, \big((1 - n_y^2 - n_z^2)^{1/2}  \partial_x \Phi  \big)
+ \zeta \, n_y\, \big( -(1 - n_y^2 - n_z^2)^{- 1/2} ( n_y \partial_x n_y + n_z \partial_x n_z)  + \partial_y n_y )
\, , \
\nonumber \\
D_{z} = \epsilon_0 \,\epsilon_\perp E_0(t) +\, \epsilon_0 \,(\epsilon_\parallel-\epsilon_\perp)  n^2_z E_0(t)
 -\,\epsilon_0 \, (\epsilon_\parallel-\epsilon_\perp)\, n_z ((1 - n_y^2 - n_z^2)^{1/2}\, \partial_x \Phi + n_y \partial_y \Phi)
  \nonumber \\
 + \,\zeta \, n_z\, \big( -(1 - n_y^2 - n_z^2)^{- 1/2} ( n_y \partial_x n_y + n_z \partial_x n_z) + \partial_y n_y  \big)
\, . \
\label{zDi}
\end{eqnarray}
Then the equation for potential $\Phi$ follows from the substitution of expressions (\ref{zDi}) in the Eq. (\ref{DinD1}).

In turn,
\begin{eqnarray}
\Xi_i =K \nabla^2 n_i
+ \epsilon_0 \,{(\epsilon_\parallel-\epsilon_\perp)}   E_i \bm n \bm E
+\zeta E_i (\nabla \bm n) -\zeta\partial_i(\bm n \bm E)  \, , \
\nonumber
\end{eqnarray}
and  for $\Xi_i^\perp = \Xi_j \delta^\perp_{ij}= \Xi_j (\delta_{ij}-n_i n_j) $ we have:
\begin{eqnarray}
\Xi_x^\perp =  \Xi_j (\delta_{xj}-n_x n_j)
=  \Xi_{x}(1 - n_x^2)- (\Xi_{y} n_x n_y) - (\Xi_{z} n_x n_z)
 \, , \
\label{pGXx}\\
\Xi_{y}^{\perp}= \Xi_j (\delta_{yj}-n_y n_j) = \Xi_{y}\, (1 - n_y^2) - \, \Xi_{x}\,(1 - n_y^2 - n_z^2)^{1/2} n_y - \Xi_{z} n_y n_z
\,\, , \
\label{pGXy}\\
\Xi_z^{\perp}= \Xi_j (\delta_{zj}-n_z n_j) = \Xi_{z}\, (1 - n_z^2) - \, \Xi_{x} \,(1 - n_y^2 - n_z^2)^{1/2} n_z - \Xi_{y} n_y n_z
\,\, . \
\label{pGXz}
\end{eqnarray}

Thus, using $\Xi_i^{\perp}$ in Eq. (\ref{trun4}), one can obtain  the general nonlinear dynamic equations for $n_y$, $n_z$ from the Eq. (\ref{trun4}):
\begin{eqnarray}
\partial_t n_{y}=
-\bm v \nabla n_y + n_k (\delta_{yj} - n_{y}n_j)\partial_k v_j +\frac{1}{\gamma} \Xi^{\perp}_{y} \, , \
\label{pGny}\\
\partial_t n_z=
-\bm v \nabla n_z + n_k \delta_{zj}^\perp \partial_k v_j +\frac{1}{\gamma} \Xi^{\perp}_z  \, . \
\label{pGnz}
\end{eqnarray}

In the next step  we can rewrite the equations (\ref{trun3}) for  the velocity components as
\begin{eqnarray}
(\rho\partial_t -\eta \nabla^2) {\bm v}
= - \nabla P + {\bm f} \,
 \ , \
\label{NvSt}
\end{eqnarray}
where  for the case of equal Frank modulus values   the expressions for the effective force $\bm f$  components have the view
\begin{eqnarray}
f_i \, =   -\nabla \left(K\nabla n_j \partial_i n_j\right) -\partial_k((\Xi_i^\perp) n_k)
 +\partial_k \left(D_k E_i\right)
\, . \
\label{Gfi}
\end{eqnarray}


Let note, that for different values of Frank modulus values $K_\alpha$ the terms $\delta \Xi_x^{(K)}$ proportional to Frank modulus $K$ to expressions for $\Xi_i $  must be replaced by the following, $\delta \Xi_x^{(K_\alpha)}$:

\begin{eqnarray}
\delta \Xi_x^{(K_\alpha)}
 =  K_3\, \Delta n_x + (K_1 - K_3)\, \big(\partial_x \big(\nabla  {\bf n} \big) \big) - (K_2 - K_3)\, \big(\partial_y n_z -  \partial_z n_y \big)\, \big({\bf n} \, \big[\nabla  {\bf n} \big] \big)
   \,
     \nonumber \\
    +\, (K_2 - K_3)\,\Big\{ \partial_z\, \Big(n_y\,  \big({\bf n} \, \big[\nabla  {\bf n} \big] \big)\Big) \,-\,  \partial_y\, \Big(n_z\,  \big({\bf n} \, \big[\nabla  {\bf n} \big] \big) \Big) \Big\}
\,\, , \
\label{delXx}
\end{eqnarray}
\begin{eqnarray}
\delta \Xi_y^{(K_\alpha)}
 =  K_3\, \Delta n_y + (K_1 - K_3)\, \big(\partial_y \big(\nabla  {\bf n} \big) \big) + (K_2 - K_3)\, \big(\partial_x n_z -  \partial_z n_x \big)\, \big({\bf n} \, \big[\nabla  {\bf n} \big] \big)
   \,
     \nonumber \\
    +\, (K_2 - K_3)\,\Big\{- \partial_z\, \Big(n_x\,  \big({\bf n} \, \big[\nabla  {\bf n} \big] \big)\Big) \,+\,  \partial_x\, \Big(n_z\,  \big({\bf n} \, \big[\nabla  {\bf n} \big] \big) \Big) \Big\}
\,\, , \
\label{delXy}
\end{eqnarray}
\begin{eqnarray}
\delta \Xi_z^{(K_\alpha)}
 =  K_3\, \Delta n_z + (K_1 - K_3)\, \big(\partial_z \big(\nabla  {\bf n} \big) \big) - (K_2 - K_3)\, \big(\partial_x n_y -  \partial_y n_x \big)\, \big({\bf n} \, \big[\nabla  {\bf n} \big] \big)
   \,
     \nonumber \\
    +\, (K_2 - K_3)\,\Big\{ \partial_y\, \Big(n_x\,  \big({\bf n} \, \big[\nabla  {\bf n} \big] \big)\Big) \,-\,  \partial_x\, \Big(n_y\,  \big({\bf n} \, \big[\nabla  {\bf n} \big] \big) \Big) \Big\}
\,\, , \
\label{delXz}
\end{eqnarray}
where
\begin{eqnarray}
 \big({\bf n} \, \big[\nabla  {\bf n} \big] \big)
 =  n_x \,  \big(\partial_y n_z -  \partial_z n_y \big) \,
  +\, n_y \,  \big(\partial_z n_x  - \partial_x n_z \big) + n_z \,  \big( \partial_x n_y  - \partial_y n_x \big)\, . \
\label{exK2}
\end{eqnarray}

Doing so,  we obtain new expressions for contributions concerned with $K_\alpha$:  $(\delta \Xi_i^\perp)^{(K_\alpha)} = \delta \Xi_j^{(K_\alpha)} \delta^\perp_{ij}= \delta \Xi_j^{(K_\alpha)} (\delta_{ij}-n_i n_j) $ to $\Xi_i^\perp $  in  nonlinear equations (\ref{pGny}), (\ref{pGnz}) for  $n_y, n_z$

\begin{eqnarray}
(\delta \Xi_x^\perp)^{(K_\alpha)}=  \delta \Xi_j^{(K_\alpha)} (\delta_{xj}-n_x n_j) =  \delta\Xi_{x}^{(K_\alpha)}(n_y^2 + n_z^2)- \delta\Xi_{y}^{(K_\alpha)}  (1 - n_y^2 - n_z^2)^{1/2} n_y) - \delta\Xi_{z}^{(K_\alpha)}  (1 - n_y^2 - n_z^2)^{1/2} n_z)\,\, , \
\label{delXxp} \\
(\delta \Xi_y^\perp)^{(K_\alpha)}= \delta \Xi_j^{(K_\alpha)} (\delta_{yj}-n_y n_j) = \delta\Xi_{y}^{(K_\alpha)}\, (1 - n_y^2) - \, \delta\Xi_{x}^{(K_\alpha)}\,(1 - n_y^2 - n_z^2)^{1/2} n_y - \delta\Xi_{z}^{(K_\alpha)} n_y n_z
\,\, , \
\label{delXyp}\\
(\delta \Xi_z^\perp)^{(K_\alpha)}= \delta \Xi_j^{(K_\alpha)} (\delta_{zj}-n_z n_j) = \delta\Xi_{z}^{(K_\alpha)}\, (1 - n_z^2) - \, \delta\Xi_{x}^{(K_\alpha)} \,(1 - n_y^2 - n_z^2)^{1/2} n_z - \delta\Xi_{y}^{(K_\alpha)} n_y n_z
\,\, . \
\label{delXzp}
\end{eqnarray}

In turn,   for different values of Frank elastic modulus values $K_\alpha$   one can obtain the following general expression  for $f_i$ components  instead of ones in  Eq. (\ref{Gfi}) from the general equation for
the momentum density \cite{PM23}:
\begin{eqnarray}
 f_i^{(K_\alpha)} =\,- \partial_x\Big\{ (K_1 - K_3)\,  \big(\nabla  {\bf n} \big)\, \partial_i n_x  \Big\}- \partial_x\Big\{ (K_2 - K_3)\, n_z\, \big({\bf n} \, \big[\nabla  {\bf n} \big] \big)\, \partial_i n_y  \Big\}
 \,
      \nonumber \\
\,- \partial_y\Big\{ (K_1 - K_3)\,  \big(\nabla  {\bf n} \big)\, \partial_i n_y  \Big\} + \partial_y\Big\{ (K_2 - K_3)\, n_z\, \big({\bf n} \, \big[\nabla  {\bf n} \big] \big)\, \partial_i n_x  \Big\}
 \,
      \nonumber \\
 \,+\,\partial_x\Big\{ (K_2 - K_3)\, n_y\, \big({\bf n} \, \big[\nabla  {\bf n} \big] \big)\, \partial_i n_z  \Big\}
  \,-\,\partial_y\Big\{ (K_2 - K_3)\, n_x\, \big({\bf n} \, \big[\nabla  {\bf n} \big] \big)\, \partial_i n_z  \Big\}
    \nonumber \\   \,
      -\nabla \left(K_3\nabla n_j \partial_i n_j\right) -\partial_k((\Xi_i^\perp)^{(K_\alpha)} n_k)
 +\partial_k \left(D_k E_i\right) \ . \
\label{dfxK}
\end{eqnarray}
Thus, for the case of different values of Frank elastic modulus values $K_\alpha$,  we obtain the following  general equations for velocity components:
\begin{eqnarray}
\rho \partial_t v_i = \eta \nabla^2 v_i
+  f_i^{(K_\alpha)}
-\partial_i P  \, , \
\label{trun3K}
\end{eqnarray}
and general equation for the pressure $P$  has the view:
\begin{equation}
 {\mathcal{C}_P} \partial_t P= -\,G_P P +  \nabla^2 P - \partial_i  f_i^{(K_\alpha)} \ . \
\label{trunpK}
\end{equation}

\end{document}